\documentclass{aa} % for a referee version
\usepackage[usenames,dvipsnames]{xcolor}
\usepackage{graphicx}
\usepackage{natbib}
\usepackage{ulem}
\usepackage{graphicx}
\usepackage{longtable}
\usepackage{subfig}
\usepackage{float}
\usepackage{txfonts}
 \newcommand{\teff}{T_\mathrm{eff}}

\begin{document}

    \title{The Effective Temperature Scale of M dwarfs}
\titlerunning{The $\teff$-scale of M dwarfs.}
\authorrunning{Rajpurohit et al.}

  \author{A. S. Rajpurohit \inst{1}, C. Reyl\'{e}\inst{1}, F. Allard\inst{2}, D. Homeier\inst{2}, M. Schultheis\inst{3}, M. S. Bessell\inst{4}, A. C.
Robin\inst{1}}

\institute{Institut UTINAM CNRS 6213, Observatoire des Sciences de l'Univers THETA Franche-Comt\'{e}-Bourgogne, Universit\'e de Franche Comt\'{e}, 
Observatoire de Besan\c{c}on, BP 1615, 25010 Besan\c{c}on Cedex, France
\and
Centre de Recherche Astrophysique de Lyon, UMR 5574: CNRS, Universit\'{e} de Lyon, \'{E}cole Normale Sup\'{e}rieure de Lyon, 46 all\'{e}e d'Italie,
69364 Lyon Cedex 07, France
\and
Universit\'{e} de Nice Sophia-Antipolis, CNRS, Observatoire de C\^{o}te d'Azur, Laboratoire Cassiop\'{e}e, 06304 Nice Cedex 4, France
\and
Research School of Astronomy and Astrophysics, Mount Stromlo Observatory, Cotter Road, Weston Creek, ACT 2611, Australia
   }
   \date{Received ...; accepted ...}

 \abstract
{Despite their large number in the Galaxy, M dwarfs remain elusive objects and the modeling of their photospheres has long remained a challenge
(molecular opacities,  dust cloud formation).}
 {Our objectives are to validate the BT-Settl model atmospheres, update the M dwarf $\teff$-spectral type relation, and find the atmospheric parameters of the stars in our sample.}
 {We compare two samples of optical spectra covering the whole M dwarf sequence with the most recent BT-Settl synthetic spectra and use a $\chi^2$
minimization technique to determine $\teff$. The first sample consists of 97 low-resolution spectra obtained with NTT at La Silla Observatory. The
second sample contains 55 mid-resolution spectra obtained at  the Siding Spring Observatory (SSO). The spectral
typing is realized by comparison with already  classified M dwarfs.}
 {We show that the BT-Settl synthetic spectra reproduce the slope of the spectral energy distribution and most of its features. Only the CaOH band at
5570\AA\ and AlH and NaH hydrides in the blue part of the spectra are still missing in the models. The $\teff$-scale obtained with the higher
resolved SSO 2.3~m
spectra is consistent with that obtained with the NTT spectra.
We compare our $\teff$-scale with those of other authors and to published isochrones using the BT-Settl colors. We also present relations between
effective temperature, spectral type and colors of the M dwarfs. }
 {} 

   \maketitle
   %
%________________________________________________________________

\section{Introduction}
Low-mass stars of less than 1~M$_{\odot}$ are the dominant stellar component of the Milky Way. They constitute 70\% of all stars \citep[][Bochanski et
al.  2010]{Reid1997}\nocite{Bochanski2010} and 40\% of the total stellar mass of the Galaxy \citep{Chabrier03,Chabrier05}. Our understanding of the
Galaxy therefore relies upon the description of this faint component. Indeed, M dwarfs have been employed in several Galactic studies as they carry
the fundamental information regarding the stellar physics, galactic structure and its formation and dynamics. Moreover, M dwarfs are now known to host
exoplanets, including super-Earth exoplanets \citep[][Udry et al. 2007]{Bonfils2007,Bonfils2012}\nocite{Udry2007}. The determination of accurate
fundamental parameters for M dwarfs has therefore relevant implications for both stellar and Galactic astronomy. Because of their intrinsic faintness
and difficulties in getting homogeneous samples with respect to age and metallicity, their physics is not yet well understood. 

Their atmosphere has been historically complex to model with the need for computed and ab initio molecular line lists accurate and complete to high
temperatures. But for over ten years already, water vapor \citep[][Barber et al. 2006]{AMESH2O}\nocite{Barber2006} and titanium oxide
\citep{Plez1998} line lists, the two most important opacities in strength and spectral coverage, have become available meeting these conditions.
And indeed, the \texttt{PHOENIX} model atmosphere synthetic spectral energy distribution improved greatly from earlier studies \citep[Hauschildt
et al. 1999]{AH95}\nocite{Hauschildt1999} to the more recent models by \cite{Allard2001,Allard2011,Allard2012} and by \cite{Witte2011} using the
most recent water vapor opacities.  

The $\teff$-scale of M dwarfs remain to this day model-dependent to some level. Many efforts have been made to derive the effective temperature
scale of M dwarfs. Due to the lack of very reliable model atmosphere, indirect methods such as blackbody fitting techniques have historically been
used to estimate the effective temperature. The \cite{Bessell1991} $\teff$-scale was based on black-body fits to the near-IR (JHKL) bands by
\cite{Pettersen1980} and \cite{Reid1984}. The much cooler black-body fits shown from \cite{Wing1979} and \cite{Veeder1974} were fits to the optical.
Their fitting line
was a continuation of the empirical $\teff$ relation for the hotter stars through the \cite{Pettersen1980} and \cite{Reid1984} IR fits for the
cooler stars. The work by \cite{Veeder1974}, \cite{Berriman1987}, \cite{Berriman1992} and \cite{Tinney1993} also used the blackbody fitting
technique to estimate the $\teff$. \cite{Tsuji1996b} provide good $\teff$ using infrared flux method (IRFM). \cite{Casagrande2008} provides a
modified IRFM $\teff$ for dwarfs including M dwarfs. These methods tend to underestimate $\teff$ since the blackbody carries little flux compared
to the M dwarfs in the Rayleigh Jeans tail redwards of $2.5~\mu$m. Temperature derived from fitting to model spectra \citep{Kirkpatrick1993} are
systematically $\sim$ 300~\,K warmer than those empirical methods. These cooler $\teff$-scale for M dwarfs was corrected recently by
\cite{Casagrande2012} bring it close to the \cite{Bessell1991,Bessell1995} $\teff$-scale.

\cite{Tinney1998} determined an M dwarf $\teff$-scale in the optical by ranking the objects in order of TiO, VO, CrH and FeH equivalent widths.
\cite{Delfosse1999} pursued a similar program in the near-infrared (hereafter NIR) with H$_2$O indices. \cite{Tokunaga1999} used a spectral color
index based on moderate dispersion spectroscopy in the K band. \cite{Leggett1996} used observed NIR low resolution spectra and photometry to
compare with the AMES-Dusty models \citep{Allard2001}. They found radii and effective temperature which are consistent with the estimates based on
photometric data from interior model or isochrone results. \cite{Leggett1998,Leggett2000} revised their results by comparing the spectral energy
distribution and NIR colors of M dwarfs to the same models. Their study provided for the first time a realistic temperature scale of M dwarfs.

In this paper, we present a new version of the BT-Settl models using the TiO line list by \cite[][and B. Plez, private communication]{Plez1998}
which is an important update since
TiO accounts for the most important features in the optical spectrum. 
Compared to the version presented in \cite{Allard2012} that was using \cite{Asplund2009} solar abundances, this new BT-Settl model uses
also the latest solar elemental abundances by \cite{Caffau2011}. We compare the revised BT-Settl synthetic spectra with the observed spectra of
152 M dwarfs using spectral synthesis and $\chi^2$ minimization techniques, and color-color diagrams to obtain the atmospheric parameters
(effective temperature, surface gravity and metallicity).
We determine the revised effective temperature scale along the entire M dwarfs spectral sequence and compare these results to those obtained by
many authors. Observations and spectral classification are presented in section~\ref{S_obs}. Details of the the model
atmospheres are described in section ~\ref{S_mod} and the $\teff$ determination is explained in section~\ref{S_teffd}. The comparison between
observations and models is done in section~\ref{S_comp} where spectral
features and photometry are compared. The effective temperature scale of M dwarfs is presented in this section. Conclusions are given in
section~\ref{S_concl}.

\section{Observations}
\label{S_obs}

We carried out spectroscopic observations on the 3.6m New Technology Telescope (NTT) at La Silla Observatory (ESO, Chile) in November 2003.
Optical low-resolution spectra were obtained in the Red Imaging and Low-dispersion spectroscopy  (RILD) observing mode with the EMMI instrument.
The spectral dispersion of the grism we used is 0.28 nm/pix, with a wavelength range 385-950 nm. We used an order blocking filter to avoid
the second order overlap that occurs beyond 800 nm. Thus the effective wavelength coverage ranges from 520 to 950 nm. The slit was 1 arcsec wide
and the resulting resolution was 1\,nm. The seeing varied from 0.5 to 1.5 arcsec. Exposure time ranged from 15~s for the brightest to 120~s
for the faintest dwarf ($I$ = 15.3). The reduction of the spectra was done using the context  \textit{long} of MIDAS. Fluxes were calibrated with
the spectrophotometric standards LTT~2415 and Feige~110.

We obtained spectra for 97 M dwarfs along the entire spectral sequence. They are presented in \cite{Reyle2006,Phan-Bao2005,Crifo2005,Martin2010}.
The list of stars, their spectral types and their optical and NIR photometry are given in Table 1. The photometry has been compiled using the
Vizier catalog access through the Centre de Donn\'{e}es astronomiques de Strasbourg. It comes from the NOMAD catalogue \citep{Zacharias2005}, the
Deep Near-Infrared Survey \citep[DENIS,][] {epchtein1997}, the Two Micron All Sky Survey \citep[2MASS,][]{Skrutskie2006},
\cite{Reid2004,Reid2007}, \cite{Koen2002,Koen2010}.\\

The observations of 55 additional M dwarfs at Siding Spring Observatory (hereafter SSO) were carried out using the Double Beam Spectrograph (DBS) that uses a dichroic
beamsplitter to separate the blue
(300--630 nm) and red (620--1000 nm) light. The blue camera with a 300 l/mm grating provided a 2 pixel resolution of 0.4 nm and the red camera
with a 316 l/mm grating provided a 2 pixel resolution of 0.37 nm. The detectors were E2V 2048x512 13.5 micron/pixel CCDs.    The observations were
taken on Mar 27 2008. The spectrophotometric standards used were HD44007, HD45282, HD55496, HD184266, and HD187111 from the STIS Next Generation
Spectral Library (NGSL, version 1)\footnote{http://archive.stsci.edu/prepds/stisngsl/index.html}, and  L745-46a and EG131 from
http://www.mso.anu.edu.au/$\sim$bessell/FTP/Spectrophotometry/. The list of stars with their photometry are given in table 2.

Spectral types for the NTT sample are obtained by visual comparison with a spectral template of comparison stars, observed together with the
targets stars at NTT as explained in \cite{Reyle2006}. For comparison, we also derive spectral types using the classification scheme based on the
TiO and CaH bandstrength \citep{Reid1997}. However no comparison stars have been observed with the DBS at
SSO. Thus spectral types for the SSO sample are computed from TiO and CaH bandstrength. Although the instrument is different, we allow to use the
comparison stars observed with EMMI on the NTT as a final check. The results agree within 0.5 subclass.

\begin{table*}[h!]
 \caption{Observable and physical quantities for our sample of stars observed at NTT with EMMI.}
\begin{tabular}{llllllllllll}
\hline
Name    &Spectral Type     &$\teff$& $\teff$ $^b$ & log\ g   &$V$ &$R$    &$I$    &$J$&    $H$&    $K$    \\
  &   & (K)  &(K)&($cm s^{-2}$)  && &&&&    \\
\hline
Gl143.1$^a$            &K7    &3900 &--- & 5.0 &10.03 &9.15    &---    &---    &---    &---\\
LHS141            &M0    &3900 &--- &5.0  &10.15    &9.35 &8.38    &7.36    &6.76    &6.58\\
LHS3833$^a$            &M0.5    &3800 &--- & 5.0 &10.06 &9.33    &---    &---    &---    &---\\
HD42581$^a$            &M1    &3700 &--- & 5.0 &8.12    &7.16 &6.12    &---    &---    &---\\
LHS14$^a$            &M1.5    &3600  &---&5.0  &10.04 &9.09    &7.99    &---    &---    &---\\
LHS65$^a$            &M1.5    &3600&3567 & 5.0  &10.86 &10.31    &10.64    &---    &---    &---\\
L127-33            &M2    &3500 &--- & 5.0 &14.19 &14.04    &12.41    &11.17    &10.58 &10.32    \\
NLTT10708        &M2    &3500 &---& 5.0  &11.16    &10.31 &9.17    &7.86    &7.28    &6.98    \\
LP831-68        &M2    &3500 &---& 5.0  &11.02    &10.02 &8.80    &7.61    &6.95    &6.69        \\
NLTT83-11        &M2    &3500&--- & 5.0  &12.90    &12.25 &11.00    &9.68    &9.01    &8.78    \\
APMPMJ0541-5349        &M2    &3500 &---& 5.0  &13.30 &12.84    &11.77    &10.64    &10.17    &9.89\\
LHS1656            &M2.5    &3400 &---& 5.0  &13.30 &12.44    &10.75    &9.52    &8.94    &8.65\\
LP763-82        &M2.5    &3400 &---& 5.0  &12.19 &11.25    &9.86    &8.55    &7.97    &7.69\\
LP849-55        &M2.5    &3400 &---& 5.0  &13.32 &13.25    &11.48    &9.97    &9.36    &9.14\\
LHS5090            &M3    &3300 &---& 5.0  &---    &14.97 &12.85    &11.58    &11.04    &10.84\\
LHS3800         &M3    &3300 &---&5.0   &---    &--- &12.23    &10.93    &10.39    &10.15    \\
LHS3842                &M3    &3300&---& 5.0   &13.80 &12.95    &11.30    &9.88    &9.29    &9.04 \\
LHS1293                &M3    &3300&---  & 5.0 &13.65 &12.66    &11.36    &9.94    &9.35    &9.07 \\
LP994-114        &M3    &3300 &---& 5.0  &---    &11.59 &10.36    &9.00    &8.37    &8.15\\
LTT9783                &M3    &3300&---&  5.0  &--- &12.11    &10.56    &9.17    &8.59    &8.34 \\
LP715-39        &M3    &3300 & 3161&5.0  &12.65    &11.53 &10.09    &8.67    &8.11    &7.82    \\
LHS1208$^a$            &M3    &3300&--- & 5.0  &9.85    &8.97 &---    &---    &---    &---\\
LEHPM4417        &M3    &3300&---&  5.0  &13.73    &13.06 &11.37    &10.09    &9.43    &9.20\\
LP831-45        &M3.5    &3200& 3125&5.0  &12.54 &11.51    &9.90    &8.49    &7.88    &7.62\\
2MASSJ04060688-0534444    &M3.5    &3200&--- & 5.0  &13.29 &12.28    &---    &9.13    &8.55    &8.30\\
LP834-32            &M3.5    &3200 &3108& 5.0  &12.38 &11.24    &9.74    &8.24    &7.65    &7.41\\
LHS502$^a$           &M3.5    &3200&---  & 5.0 &11.49 &10.43    &9.11    &---    &---    &---\\
LEHPM 1175        &M3.5    &3200&--- & 5.0  &--- &13.08    &11.51    &10.01    &9.47    &9.17\\
LEHPM1839        &M3.5    &3200&---  & 5.0 &---    &13.32 &12.11    &10.55    &9.95    &9.71\\
L130-37            &M3.5    &3200 &--- &5.0  &13.04 &11.97    &10.37    &8.94    &8.34    &8.01\\
LEHPM6577        &M3.5    &3200&--- & 5.0  &---    &13.03 &11.79    &10.34    &9.73    &9.47\\
L225-57                &M4    &3200&--- &  5.0 &--- &11.70    &9.79    &8.23    &7.61    &7.31    \\
LP942-107            &M4    &3200 &3052& 5.0  &13.93 &12.73    &11.13    &9.63    &9.08    &8.77 \\
LP772-8                &M4    &3200&--- & 5.0  &14.11 &13.43    &11.52    &10.05    &9.48    &9.20\\
LP1033-31            &M4    &3200&---  & 5.0 &--- &12.12    &10.54    &9.10    &8.46    &8.21  \\
L166-3                &M4    &3200&--- & 5.0  &--- &12.76    &11.33    &9.83    &9.28    &9.00  \\
LP877-72            &M4    &3200&---  &5.0  &--- &11.---    &10.22    &8.86    &8.24    &8.00     \\
LP878-73            &M4    &3200&--- & 5.0  &14.55 &14.22    &12.63    &10.86    &10.27    &10.00     \\
LP987-47            &M4    &3200&--- &5.0   &---    &--- &10.82    &9.41    &8.78    &8.55      \\
LP832-7                &M4    &3200&--- & 5.0  &14.09 &13.45    &---    &9.87    &9.24    &8.98 \\
LHS183            &M4    &3200&--- & 5.0  &12.79 &11.51    &---    &8.57    &8.00    &7.75  \\
LHS1471            &M4    &3200 &---&  5.0 &---    &13.22 &11.56    &9.94    &9.37    &9.08\\
APMPMJ2101-4125        &M4    &3200&--- & 5.0  &--- &13.34    &11.47    &9.96    &9.38    &9.09\\
APMPMJ2101-4907        &M4    &3200&--- & 5.0  &--- &---    &10.52    &9.12    &8.48    &8.19\\
LEHPM3260        &M4    &3200&--- & 5.0  &---    &12.53 &10.60    &9.13    &8.54    &8.19\\
LEHPM3866        &M4    &3200&---  & 5.0 &---    &--- &11.82    &10.21    &9.58    &9.29\\
LEHPM5810        &M4    &3200&--- & 5.0  &---    &13.58 &11.66    &9.91    &9.33    &9.05\\
LHS5045                &M4.5    &3100&--- & 5.0  &--- &---    &10.78    &9.17    &8.60    &8.24\\
LP940-20            &M4.5    &3100 &--- & 5.0 &--- &14.87    &12.65    &10.92    &10.32    &10.01\\
L170-14A            &M4.5    &3100&--- & 5.0  &--- &12.86    &11.50    &9.76    &9.13    &8.88\\
LHS1201                &M4.5    &3100&--- &  5.0 &17.55 &15.52    &12.90    &11.12    &10.52    &10.25\\
LHS1524                &M4.5    &3100&---  &5.0  &--- &14.45    &12.65    &10.98    &10.45    &10.17\\
LTT1732                &M4.5    &3100&--- & 5.0  &--- &13.19    &11.27    &9.69    &9.11    &8.80\\
LP889-37            &M4.5    &3100& 2923 & 5.0  &14.52 &13.21    &11.46    &9.77    &9.16    &8.82\\
LHS5094                &M4.5    &3100&--- & 5.0  &14.02 &12.72    &10.97    &9.30    &8.72    &8.41\\
LP655-43            &M4.5    &3100 &2924&5.0   &14.44 &13.14    &11.41    &9.73    &9.14    &8.82\\
LHS138$^a$             &M4.5    &3100&--- & 5.0  &12.07 &10.70    &8.94    &---    &---    &---\\
APMPMJ1932-4834        &M4.5    &3100 &---& 5.0  &--- &14.38    &12.37    &10.63    &10.02    &9.72\\
2MASSJ23522756-3609128    &M4.5    &3100 &--- & 5.0 &--- &17.27    &---    &13.09    &12.57    &12.28\\
LEHPM640        &M4.5    &3100 &---& 5.0  &17.74 &14.26    &12.30    &10.76    &10.14    &9.90\\
LEHPM1853        &M4.5    &3100 &--- & 5.0 &---    &12.77 &11.03    &9.46    &8.85    &8.61\\
LEHPM3115        &M4.5    &3100 &--- & 5.0 &---    &13.94 &12.10    &10.49    &9.92    &9.63\\
LEHPM4771        &M4.5    &3100 &--- & 5.0 &17.74 &13.79    &11.29    &9.54    &8.95    &8.63\\
LEHPM4861        &M4.5    &3100 &--- & 5.0 &---    &13.28 &11.75    &10.13    &9.60    &9.34\\
L291-115            &M5    &2900&--- & 5.0  &15.88 &14.90    &12.26    &10.44    &9.83    &9.54     \\
LP904-51        &M5    &2900&--- & 5.0  &---    &15.32 &12.84    &11.04    &10.44    &10.16        \\
LHS168            &M5    &2900&---  & 5.0 &13.78 &12.60    &---    &8.77    &8.21    &7.83    \\
                 \hline
\end{tabular}
\end{table*}

\begin{table*}[t]
{\bf \small Table 1.} Continued.\vspace*{0.15cm}\\
\begin{tabular}{llllllllllll}
\hline
Name    &Spectral Type     &$\teff$& $\teff$ $^b$ & log\ g   &$V$ &$R$    &$I$    &$J$&    $H$&    $K$    \\
  &   &(K)  &(K)&($cm s^{-2}$)  && &&&&    \\
\hline
LP829-41            &M5.5    &2800&--- & 5.0  &16.10 &15.95    &13.21    &11.31    &10.76 &10.40    \\
LP941-57            &M5.5    &2800&--- & 5.0  &--- &14.88    &12.98    &11.06    &10.47 &10.13    \\
LHS546            &M5.5    &2800&---   & 5.0&14.69 &---    &---    &9.15    &8.50    &8.18\\
LP714-37        &M5.5    &2800&--- & 5.5  &16.26 &15.02    &12.99    &11.01    &10.37    &9.92 \\
LHS1326            &M6    &2800&--- & 5.5  &15.61 &14.49    &---    &9.84    &9.25    &8.93    \\
2MASSJ12363959-1722170    &M6    &2800&---  &5.0  &17.56 &15.86    &13.91    &11.67    &11.09    &10.71\\
2MASSJ21481595-1401059    &M6.5    &2700&--- & 5.0  &--- &20.20    &17.15    &14.68    &14.11    &13.65\\
2MASSJ05181131-3101519    &M6.5    &2700&---  &5.0  &17.74 &16.85    &14.17    &11.88    &11.23    &10.90\\
LP788-1            &M6.5    &2700 &---& 5.0  &--- &16.66    &13.36    &11.07    &10.47    &10.07 \\
APMPMJ1251-2121        &M6.5    &2700&--- & 5.0  &--- &16.65    &13.78    &11.16    &10.55    &10.13\\
APMPMJ2330-4737        &M7    &2700&---  &5.0  &--- &---    &13.70    &11.23    &10.64    &10.28 \\
LP789-23        &M7    &2700&--- & 5.0  &---    &17.90 &14.55    &12.04    &11.39    &10.99\\
LHS292            &M7    &2700&--- &  5.5 &15.60 &14.40    &11.25    &8.86    &8.26    &7.93 \\
2MASSJ03144011-0450316    &M7.5    &2600&--- &5.0   &--- &19.43    &---    &12.64    &12.00    &11.60 \\
LHS1604        &M7.5    &2600&--- & 5.0  &18.02    &16.52 &13.75    &11.30    &10.61    &10.23 \\
LP714-37        &M7.5    &2600&--- & 5.5  &16.26 &15.52    &12.99    &11.01    &10.37    &9.92 \\
LP655-48        &M7.5    &2600 &2250& 5.0  &17.86 &15.95    &13.35    &10.66    &9.99    &9.54  \\
LP851-346        &M7.5    &2600&--- & 5.5  &---    &16.79 &13.77    &10.93    &10.29    &9.88 \\
LHS1367                &M8    &2600&--- &  5.0 &--- &17.34    &14.18    &11.62    &10.95    &10.54 \\
2MASSJ05022640-0453583    &M8    &2600&---&  5.0  &--- &20.39    &17.35    &14.52    &13.95    &13.58 \\
LHS132            &M8    &2600&--- &  5.0 &---    &17.14 &13.83    &11.13    &10.48    &10.07       \\
2MASSJ22062280-2047058    &M8    &2600&--- &  5.0 &--- &18.93    &15.09    &12.37    &11.69    &11.31     \\
2MASSJ22264440-7503425    &M8    &2600&--- & 5.0  &--- &18.95    &15.20    &12.35    &11.70    &11.25     \\
2MASSJ04103617-1459269    &M8.5    &2500&--- & 5.5  &--- &---    &16.68    &13.94    &13.24    &12.81\\
2MASSJ05084947-1647167    &M8.5    &2500&--- & 5.5  &--- &---    &16.46    &13.69    &12.96    &12.53  \\
2MASSJ04362788-4114465    &M8.5    &2500&--- & 5.5  &--- &19.96    &16.04    &13.10    &12.43    &12.05 \\
2MASSJ10481463-3956062    &M9        &2500&--- &5.5   &--- &15.93    &12.67    &9.54    &8.90    &8.45     \\
2MASSJ20450238-6332066    &M9.5    &2500&--- & 5.5  &--- &19.24    &16.05    &12.62    &11.81    &11.21\\
2MASSJ09532126-1014205  &M9.5    &2500&--- & 5.5  &--- &19.58    &16.82    &13.47    &12.64    &12.14     \\
\hline
\end{tabular}

$^a$ Saturation in NIR bands.\\
$^b$ $\teff$ from \cite{Casagrande2008}.
\end{table*}

\begin{table*}[h!]
 \caption{Observable and physical quantities for our sample of stars observed at SSO.}
\begin{tabular}{llllllllllll}
\hline
Name    &Spectral Type     &$\teff$& $\teff$ $^b$ & log\ g   &$V$ &$R$    &$I$    &$J$&    $H$&    $K$    \\
  &   & (K)  &(K)&($cm s^{-2}$)  && &&&&    \\
\hline
HIP49986  &  M1.5  &  3700&3445  &5.0 & 9.07  &  8.21 &  7.08  &  5.89  &  5.26  &  5.01    \\
HIP82256  &  M1.5  &  3700&3470 &5.0 & 11.38  &  10.39 &  9.24  &  8.04  &  7.48  &  7.22    \\
HIP56528  &  M1.5  &  3600 &3472 &5.0  &9.81  &  8.85 &  7.66  &  6.47  &  5.86  &  5.62    \\
NLTT19190  &  M1.5  &  3600&3456  &5.0  &11.49  &  10.57 &  9.34  &  8.11  &  7.47  &  7.20    \\
NLTT42523  &  M2  &  3600 &3444 & 5.0 &12.08  &  11.06 &  9.81  &  8.60  &  8.01  &  7.80    \\
HIP80229  &  M2  &  3600 &3486 & 5.0 &11.91  &  10.90 &  9.65  &  8.48  &  7.87  &  7.64    \\
LP725-25  &  M2  &  3600 &3476 &5.0  &11.76  &  10.82 &  9.59  &  8.36  &  7.68  &  7.44    \\
HIP61413  &  M2  &  3500 &3454 &5.0 & 11.49  &  10.48 &  9.17  &  7.99  &  7.37  &  7.15    \\
LP853-34  &  M2  &  3500 & 3339&5.0  &12.32  &  11.31 &  9.99  &  8.69  &  8.10  &  7.83    \\
LP859-11  & M2  &  3500  &3433 	& 5.0 &12.00  &  10.97  & 9.69  &  8.49  &  7.88  &  7.63    \\
LP788-49  &  M2  &  3500 &3356 & 5.0 &11.81  &  10.85 &  9.55  &  8.30  &  7.74  &  7.49    \\
HIP42762  &  M2  &  3500 &3302 & 5.0 &11.75  &  10.76 &  9.42  &  8.12  &  7.49  &  7.28    \\
HIP51317  &  M2  &  3500 &3403 &5.0  &9.67  &  8.67  & 7.34  &  6.18  &  5.60  &  5.31 \\
HIP60559  &  M2  &  3500 &3382 & 5.0 &11.30  &  10.29 &  8.99  &  7.73  &  7.25  &  6.95    \\
HIP47103  &  M2  &  3500 &3319 &5.0  &10.87  &  9.89  & 8.58  &  7.34  &  6.74  &  6.47    \\
HIP93206  &  M2.5  &  3500 &3366 &5.0 & 11.23  &  10.18 &  8.80  &  7.52  &  6.93  &  6.70    \\
LP834-3  &  M2.5  &  3500 &--- & 5.0 &---  &  --- &---  &---  &---  &---    \\
HIP84521  &  M2.5  &  3500 &3345 &5.0  &11.57  &  10.53 &  9.22  &  7.93  &  7.39  &  7.11    \\
HIP91430  &  M2.5  &  3500 &3352 & 5.0 &11.32  &  10.26 &  8.92  &  7.66  &  7.06  &  6.85    \\
HIP50341  &  M2.5  &  3500&3314  & 5.0 &11.02  &  10.01 &  8.62  &  7.32  &  6.71  &  6.45    \\
LP672-2  &  M2.5  &  3400 &--- &5.0  &12.58  &  11.54 &  10.12  &  8.80  &  8.14  &  7.93    \\
NLTT24892  &  M2.5  &  3400 &3244 &5.0 & 12.52  &  11.47 &  10.05  &  8.73  &  8.118  &  7.84    \\
NLTT34577  &  M2.5  &  3400 & 3254&5.0  &12.44  &  11.40 &  9.99  &  8.64  &  8.00  &  7.80    \\
LP670-17  &  M3  &  3400 &3226 & 5.0 &12.14  &  11.08 &  9.63  &  8.28  &  7.68  &  7.39    \\
HIP59406  &  M3  &  3400 &3226 & 5.0 &11.75  &  10.69  & 9.25  &  7.89  &  7.36  &  7.04    \\
HIP74190  &  M3  &  3400 &3258 &5.0  &11.55  &  10.48 &  9.05  &  7.72  &  7.13  &  6.86    \\
NLTT46868  &  M3.5   & 3400 &3221 &5.0 & 12.23  &  11.08 &  9.61  &  8.26  &  7.73  &  7.44    \\
HIP62452  &  M4  &  3300 & 3095 & 5.0 &11.46  &  10.31 &  8.71  &  7.19  &  6.67  &  6.36    \\
NLTT25488  & M4  &  3200 & 2986& 5.0 &15.66  &  14.46 &  12.73  &  11.09  &  10.52  &  10.21    \\
NLTT29087  &  M4  &  3200&2971  &5.0 & 14.79  &  13.57 &  11.84  &  10.22  &  9.62  &  9.35    \\
NLTT29790  & M4  &  3200 &	2987 & 5.0 &14.73  &  13.54 &  11.85  &  10.22  &  9.64  &  9.34    \\
LP734-32  &  M4  &  3200 &3024 & 5.0 &12.15  &  10.99 &  9.35  &  7.77  &  7.14  &  6.86    \\
LP739-2  &  M4  &  3100 &2939 &5.0  &14.44  &  13.18 &  11.40  &  9.73  &  9.17  &  8.89    \\
LP735-29  &  M4  &  3100&2940  &5.0 & 14.18  &  12.95 &  11.18  &  9.52  &  8.97  &  8.67    \\
GJ1123  &  M4  &  3100&--- & 5.0& 13.14  &  11.90  & 10.10  &  8.33  &  7.77  &  7.45    \\
GJ1128  &  M4  &  3100 & ---& 5.0 &12.66  &  11.40  & 9.61  &  7.95  &  7.38  &  7.04    \\
NLTT35266  &  M4.5  &  3100 & 2942& 5.0& 15.15  &  13.88 &  12.05  &  10.41  &  9.94  &  9.66    \\
NLTT41951  &  M4.5  &  3100 & & 5.0 &15.06  &  13.77 &  11.99  &  10.36  &  9.80  &  9.51    \\
NLTT21329  & M4.5&3000&2949  &5.0 & 13.75  &  12.38  & 10.42  &  8.60  &  8.07  &  7.73  &      \\
LP732-35  &  M5  &  3100 &2901 &  5.0&14.10  &  12.78 &  10.94  &  9.36  &  8.76  &  8.49    \\
NLTT18930  &  M5  &  3100 &2903 & 5.0 &15.34  &  13.93 &  12.03  &  10.31  &  9.76  &  9.44    \\
2MASS J14221943-7023371 &   M5  &  3000&--- & 5.0&  ---  &  --- &---  &---  &--- &--- \\
NLTT22503  &  M5  &  3000&	2785  &5.0  &13.66  &  12.32 &  10.39  &  8.50  &  7.92  &  7.60    \\
NLTT28797  &  M5  &  3000 &2826 & 5.0 &15.62  &  14.24 &  12.32  &  10.54  &  9.99  &  9.64    \\
NLTT30693  &  M5.5  &  3000 &2785 &5.5  &15.32  &  13.86 &  11.85  &  9.95  &  9.36  &  9.00    \\
LHS288  &  M5.5  &  3000&2770  &5.0  &13.87  &  12.42 &  10.31  &  8.48  &  8.05  &  7.73    \\
GJ551  & M5.5& 2900 &--- & 5.0 &3.63  &  2.08  & 5.36  &  4.83  &  4.38  &---      \\
LHS2502  &  M6  &  2900 & 2468&5.5  &19.36  &  17.54 &  15.33  &  12.75  &  12.07  &  11.79    \\
NLTT20726  &  M6.5  &  2800& 2464 & 5.0 &16.11  &  14.24 &  11.85  &  9.44  &  8.84  &  8.44    \\
GJ406  &  M6.5  &  2800&---  & 5.5 &13.57  &  11.81 &  9.51  &  7.08  &  6.48  &  6.08    \\
LHS2351  &  M7  &  2800& 2346 &5.5 & 19.22  &  17.39 &  14.91  &  12.33  &  11.72  &  11.33    \\
SCR J1546-5534  &  M7.5  &  2700 &---&5.5 &  ---  &---  &---  &---  &---  &---    \\
GJ752b  &  M8  &  2700 & ---& 5.5& 5.01  &---  &--- &  9.91  &  9.23  &  8.76    \\
GJ644c  &  M7  &  2700 &--- & 5.5 &16.90  &  14.78  & 12.24  &  9.78  &  9.20  &  8.82   \\
LHS2397a  &  M8  &  2700 &---& 5.5 &19.66  &  17.42 &  14.86  &  11.93  &  11.23  &  10.73   \\

\hline
$^b$ $\teff$ from \cite{Casagrande2008}.
\end{tabular}

\end{table*}

\section{Model atmospheres}
\label{S_mod}

For this paper, we use the most recent BT-Settl models partially published in a review by \cite{Allard2012} and described by \cite{Allard2012b}.
These model atmospheres are computed with the \texttt{PHOENIX} multi-purpose atmosphere code version 15.5 \citep[][Allard et al.
2001]{Phoenix97}\nocite{Allard2001} solving the radiative transfer in 1D spherical symmetry, with the classical assumptions: hydrostatic
equilibrium, convection using the Mixing Length Theory, chemical equilibrium, and a sampling treatment of the opacities.  The models use
a mixing length as derived by the Radiation HydroDynamic (hereafter RHD) simulations of \cite{Ludwig2002,Ludwig2006} and \cite{Freytag2012} and a
radius as determined by the \cite{BCAH98} interior models as a function of the atmospheric parameters ($\teff$, log $g$, [M/H]).  
The BT-Settl grid extends from $\teff = 300 - 7000$\,K, log$g = 2.5 - 5.5$ and [M/H]$= -2.5 - 0.0$ accounting for alpha element enrichment.  The
reference solar elemental abundances used in this version of the BT-Settl models are those defined by \cite{Caffau2011}. The synthetic colors and
spectra are distributed with a spectral resolution of around R=100000 via the \texttt{PHOENIX} web
simulator\footnote{http://phoenix.ens-lyon.fr/simulator}.

\label{S_teffd}

\begin{figure*}
   \centering
   \includegraphics[width=9.0cm,height=12cm]{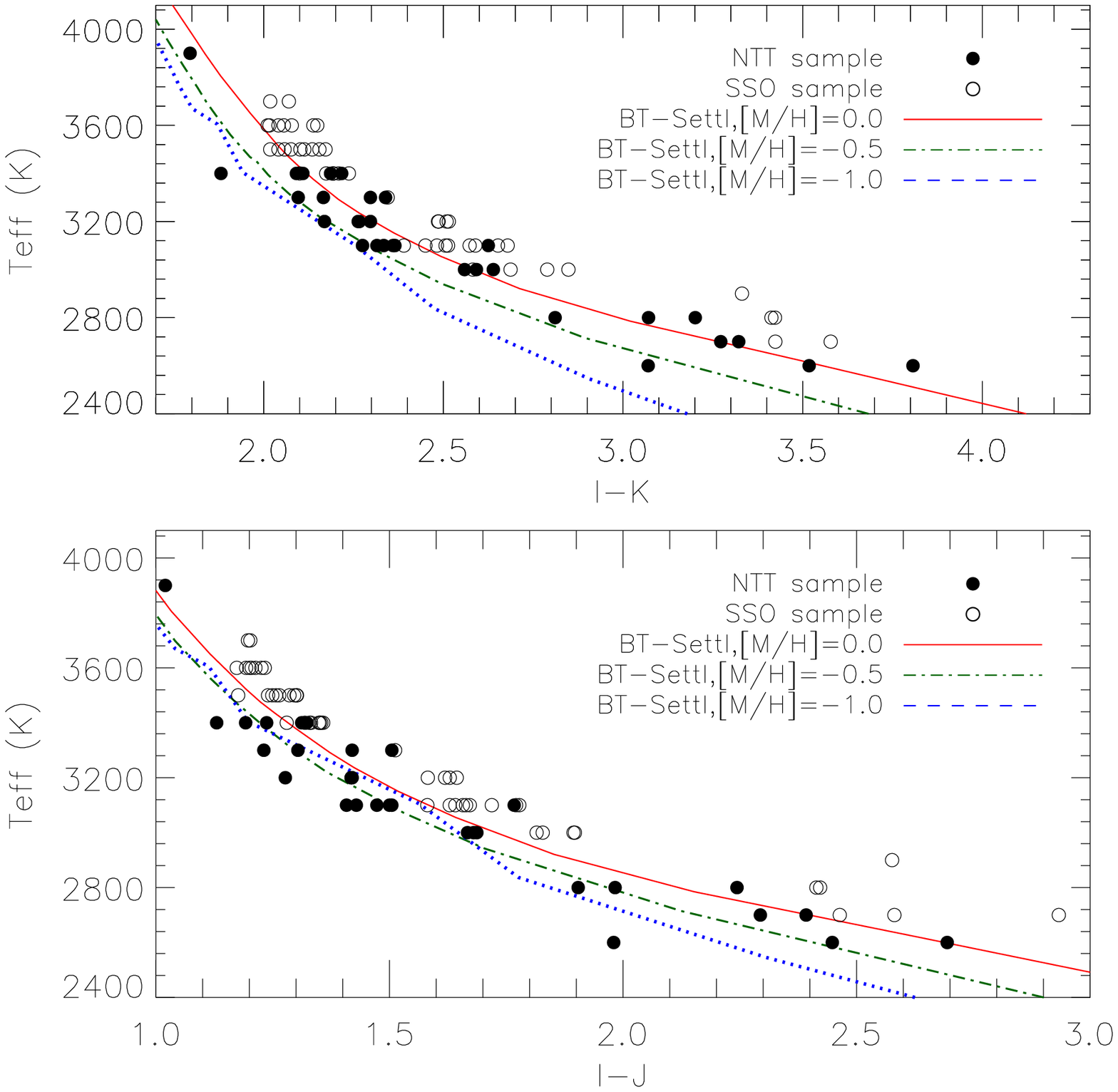}
   \includegraphics[width=9.0cm,height=12cm]{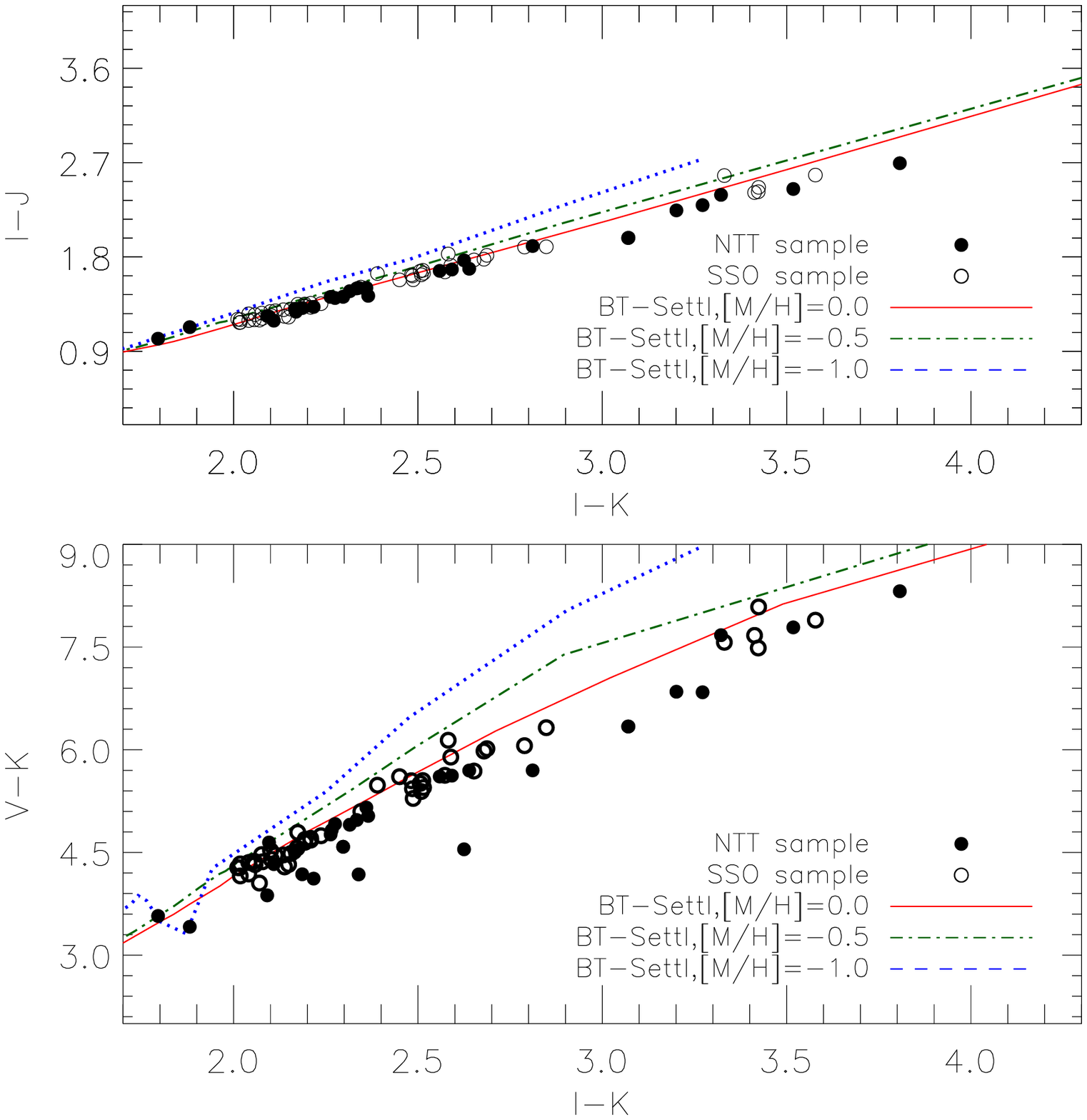}
\caption{$\teff$ vs near-infrared colors (left panel) and color-color plot (right panel) for observed M dwarfs (open and filled circle) compared
to the values obtained with the 5 Gyrs isochrones from \cite{BCAH98} at various metallicities.}
 \label{Fig:1}
   \end{figure*}

Hot temperature grains have been shown to form in the uppermost layers of M dwarfs with effective temperatures below 3000\,K, but clear effects
observable at the spectral resolution considered in this paper are only apparent below 2600\,K i.e. for later spectral type than those considered in
this paper. These grains produce a "veiling" by dust scattering over the optical band of the latest type M dwarfs.  The BT-Settl models use therefore
a slightly revised version of the \cite{Rossow1978} cloud model. See \cite{Allard2012,Allard2012c}, \cite{Allard2012b} and \cite{Rajpurohit2012a} for
details on the model construction.

Relative to previous models by \cite{Allard2001}, the current version of the BT-Settl model atmosphere is using the BT2 water vapor line list computed
by \cite{Barber2006}, TiO, VO, CaH line lists by \cite{Plez1998}, MgH by \cite[Story et al. 2003]{Weck2003}\nocite{Story2003}, FeH and CrH by
\cite[Chowdhury et al. 2006]{Dulick2003}\nocite{Chowdhury2006}, NH$_3$ by \cite{Yurchenko2011}, CO$_2$ \citep{Tashkun2004}, and H$_2$ Collision
Induced Absorption (CIA) by \cite{Borysow2001} and \cite{Abel2011}, to mention the most important.  We use the CO line list by
\cite{Goorvitch94a,Goorvitch94b}. Detailed profiles for the alkali lines are also used \citep{AllardN2007}.

In general, the \cite{Unsold1968} approximation is used for the atomic damping constants with a correction factor to the widths of 2.5 for the
non-hydrogenic atoms \citep{Valenti1996}. More accurate broadening data for neutral hydrogen collisions by \cite{Barklem2000} have been included for
several important atomic transitions such as the alkali, Ca\,I and Ca\,II resonance lines. For molecular lines, we have adopted average values
(e.\,g.\ $\langle\gamma_6^{HIT}(T_0, P_0\footnote{Standard temperature  296\,K and pressure 1 atm})\rangle_{H_2O} = 0.08 \,\, P_{\rm gas} \,
[\mathrm{cm}^{-1}\mathrm{atm}^{-1}]$ for water vapor lines) from the HITRAN database \citep{HITRAN2008},
which are scaled to the local gas pressure and temperature 
\begin{equation}
\gamma_6(T) = \langle\gamma_6^{HIT}(T_0,P_0)\rangle \left(\frac{296\,K}{T}\right)^{0.5}\, \left(\frac{P}{1\, {\rm atm}} \right) \enspace,
%\enspace. 
\end{equation} 
with a single temperature exponent of 0.5, to be compared to values ranging mainly from 0.3 to 0.6 for water transitions studied by
\citet{Gamache1996}. The HITRAN database gives widths for broadening in air, but \citet{VSTAR2012} find that these agree in general within 10\,--\,20\% with
those for broadening by a solar composition hydrogen-helium mixture.

\section{$\teff$ determination}
\label{S_teffd}

We use a  least-square minimization program using the new BT-Settl model atmospheres  
to derive a revised effective temperature  scale of M dwarfs.  The stars in our samples most probably belong to the thin disc of our Galaxy
\citep{Reyle2002,Reyle2004}. Thus we determine the $\teff$ of our targets assuming solar metallicity. This is a reasonable assumption as can be
seen in Fig.~\ref{Fig:1} where we compare our two samples to three 5 Gyrs isochrones with solar, [M/H]=~-0.5 and -1.0~dex. The samples are clearly
compatible with solar metallicity.

Both theory and observation indicate that M dwarfs have log\,$g~=~5.0 \pm 0.2$ \citep{Gizis1996,Casagrande2008} except for the latest-type M
dwarfs. We therefore restrict our analysis to log\,$g~=~5.0 - 5.5$ models.
Each synthetic spectrum was convolved to the observed spectral resolution  \and a scaling factor is applied to normalize the average flux to unity.
We then compare each of the observed spectra with all the synthetic spectra in the grid by taking the difference between the
flux values of the synthetic and observed spectra at each wavelength point. We interpolated the model spectra on the wavelength grid of the observed spectra.  The sum of the squares of these differences is obtained for each model
in the grid, and the best model for each object is selected. The best models were finally inspected visually by comparing them with the
corresponding observed spectra. 
Due to the lower signal-to-noise ratio in the SSO 2.3~m spectra bluewards of
500~nm   (see Fig.~3), especially for spectral types later than M4,  we have excluded this region  below 500~nm from the $\chi^2$ computation.
We have also checked the variation in effective temperature of the best fit as a function of the spectral type of the observed dwarfs.  We found
generally good agreement and conclude that our model fitting procedure can be used to estimate the effective temperature with an uncertainty of
$\sim$100\,K. The purpose of this fit is to determine the effective temperature by fitting the overall shape of the optical spectra. 
No attempt has been made to fit the individual atomic lines such as the K\,I and Na\,I resonance doublets. With the available resolution we
cannot constrain the metallicity; high resolution spectra would be necessary (Rajupurohit et al. in prep.). In addition, we checked the influence of the spectral resolution to our derived temperatures. We
degraded the resolution of the spectra of SSO 2.3~m down to 1~nm and redid the procedure. No systematic difference in $\teff$ was found.
The results are summarized in Table~1 and ~2.

\section{Comparison between models and observations}
\label{S_comp}

\subsection{Spectroscopic confrontation}

The optical spectrum of M dwarfs is dominated by molecular band absorption, leaving no window onto the continuum \citep{AllardPhDT90}. The major
opacity sources in the optical regions are due to  titanium oxide (TiO) and vanadium oxide (VO) bands, as well as to MgH, CaH, FeH hydrides
bands and CAOH hydroxide bands in late-type M dwarfs.  In M dwarfs of spectral type later than M6, the outermost atmospheric layers fall below the
condensation temperature of silicates, giving rise to the formation of dust clouds \citep[][Allard et al.
1997]{Tsuji1996b,Tsuji1996a}\nocite{Allard1997}.

\begin{figure*}[ht!]
   \centering
   \includegraphics[width=18cm,height=18cm]{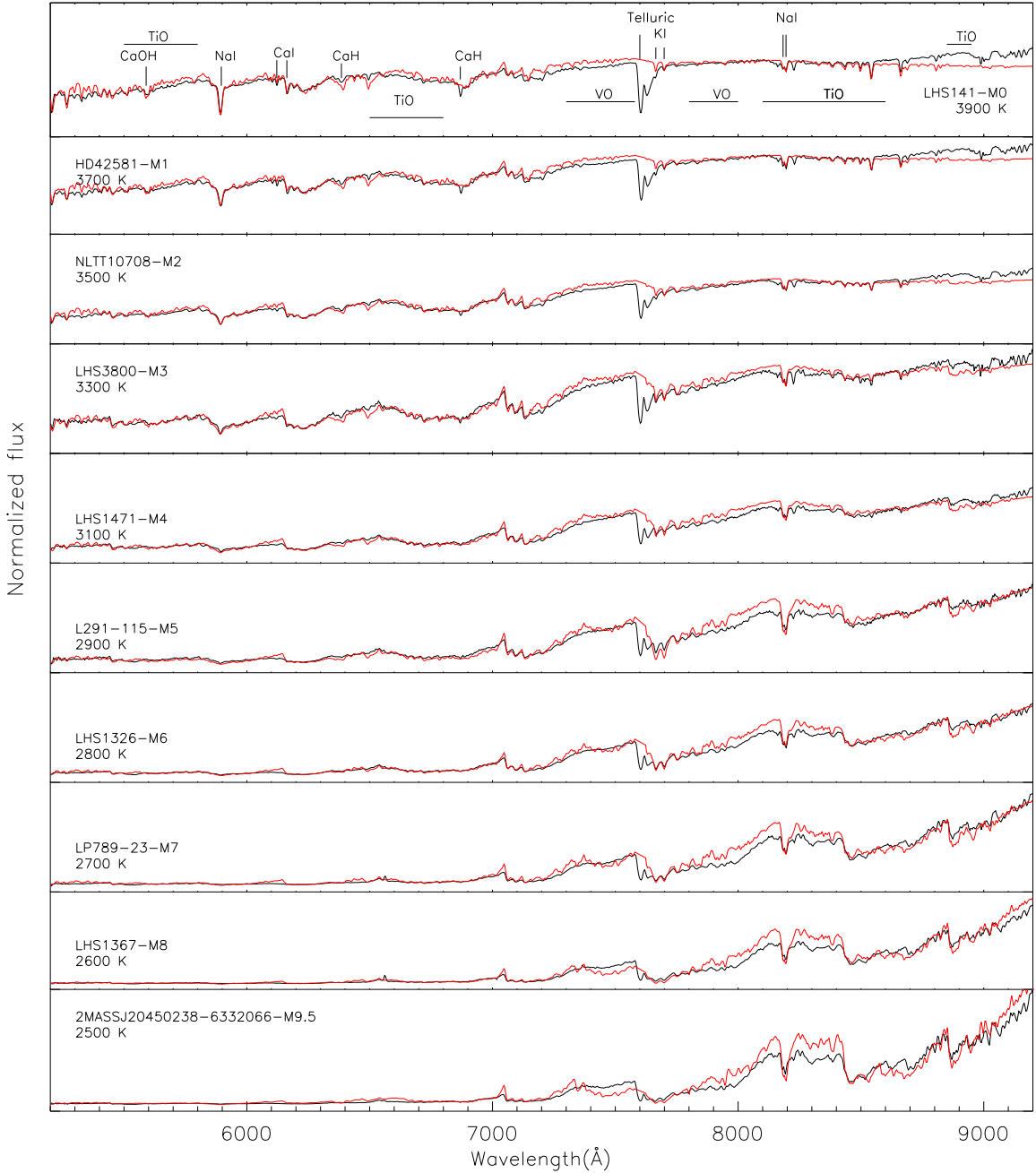}
    \caption{Optical to red SED of M dwarfs from M0 to M9.5 observed with the NTT at a spectral resolution of 10.4~\AA\ compared to the best fit
             BT-Settl synthetic spectra (red lines). The models displayed have a surface gravity of log $g$ = 5.0 to 5.5. Telluric features near
             7600\,\AA\ have been ignored from the chi-square minimization.
     }
         \label{Fig:2}
   \end{figure*}

\begin{figure*}[ht!]
   \centering
   \includegraphics[width=18cm,height=18cm]{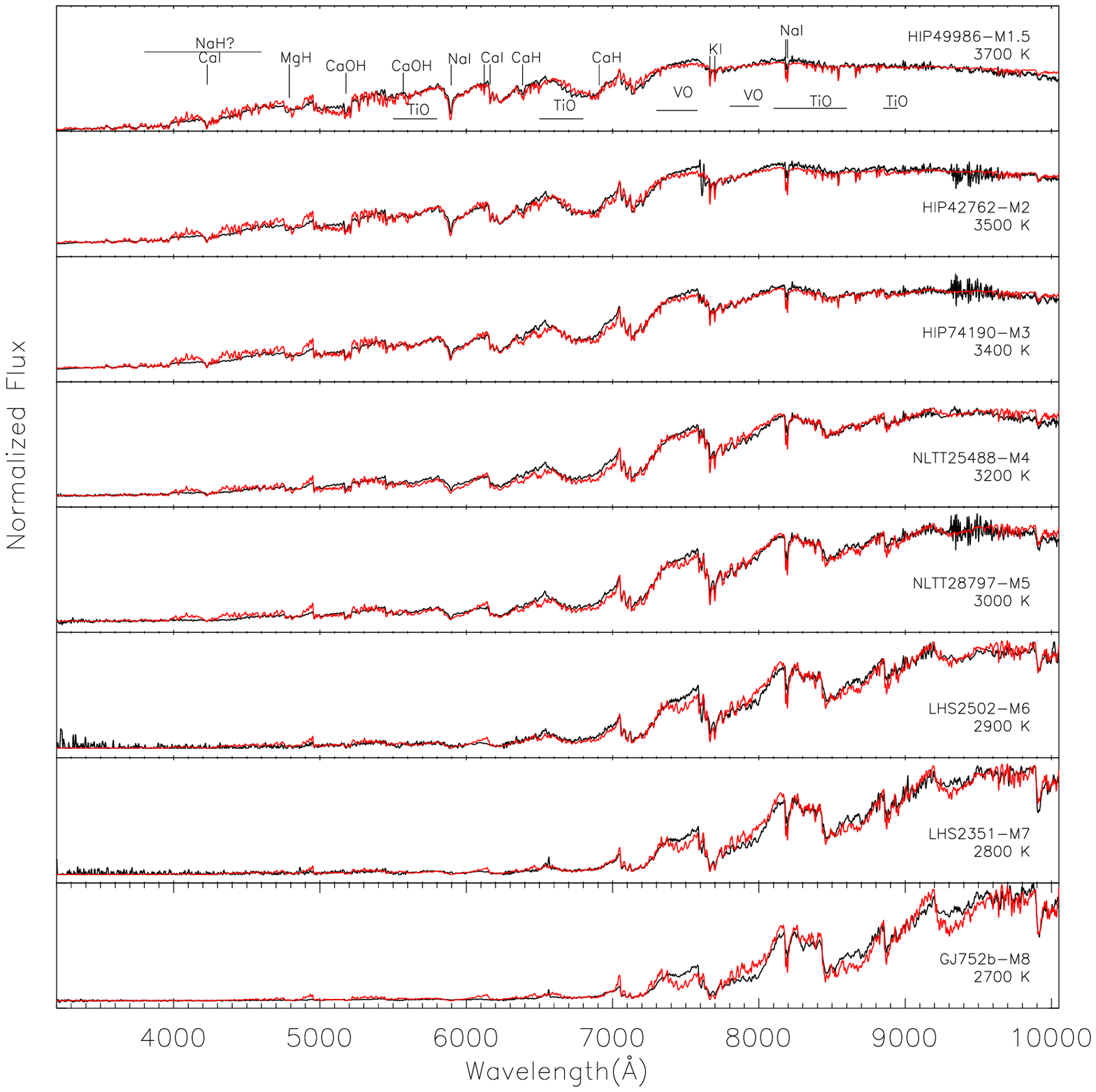}
    \caption{Optical to red SED of M dwarfs from M1 to M8 observed with the SSO 2.3~m at a spectral resolution of 1.4~\AA\
     compared to the best fitting (chi-square minimization) BT-Settl synthetic spectra (red lines). The models displayed have a surface
     gravity of log $g$=5.0 to 5.5. At blue wavelengths ($<$ 5000$\AA$) the instrumentals noise dominate the late-type M dwarfs. %is shown.
}
         \label{Fig:3}
   \end{figure*}

We compared the two samples of M dwarfs with the most recent BT-Settl synthetic spectra in Fig.~\ref{Fig:2} and \ref{Fig:3} through the entire
M dwarf spectral sequence. The synthetic spectra reproduce very well the slope of the observed spectra across the M dwarfs regime. This is a
drastic improvement compared to previous comparisons of earlier models \citep[e.g.][]{Leggett1998}. 

However, some indications of missing opacities 
persist in the blue part of the late-type M dwarf such as the B' $^{ 2}${$\Sigma$}$^{+}$$<$-- X $^{ 2}${$\Sigma$}$^{+}$
system of MgH \citep{Story2003}, as well as TiO and VO opacities around 8200 \AA. Opacities are totally
missing for the CaOH band at 5570\,\AA. The missing hydride bands of AlH and NaH between  3800 and 4600~\AA\ among others could be responsible for
the remaining discrepancies.  Note that chromospheric emission fills the Na\,I\,D transitions in the latest-type M dwarfs displayed here.
    
We see in this spectral regime no signs of dust scattering or of the weakening of features due to sedimentation onto grains until the M8 and later
spectral types where the spectrum becomes flat due to the sedimentation of TiO and VO bands and to the veiling by dust scattering.

\begin{figure*}[ht!]
 \centering
   \subfloat{\label{fig:4}\includegraphics[width=12.5cm,height=10cm]{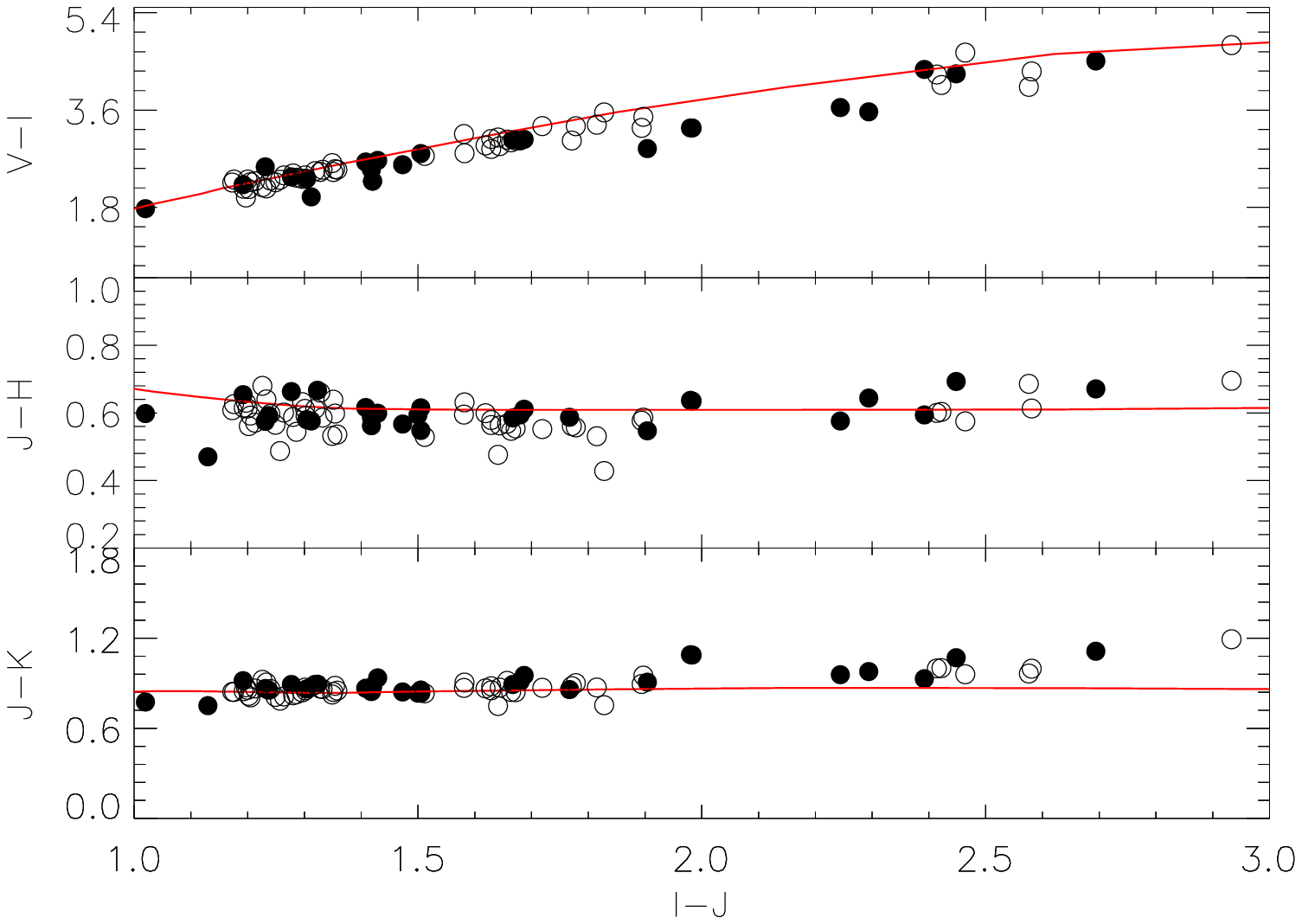}}
\qquad
   \subfloat{\label{fig:4}\includegraphics[width=12.5cm,height=10cm]{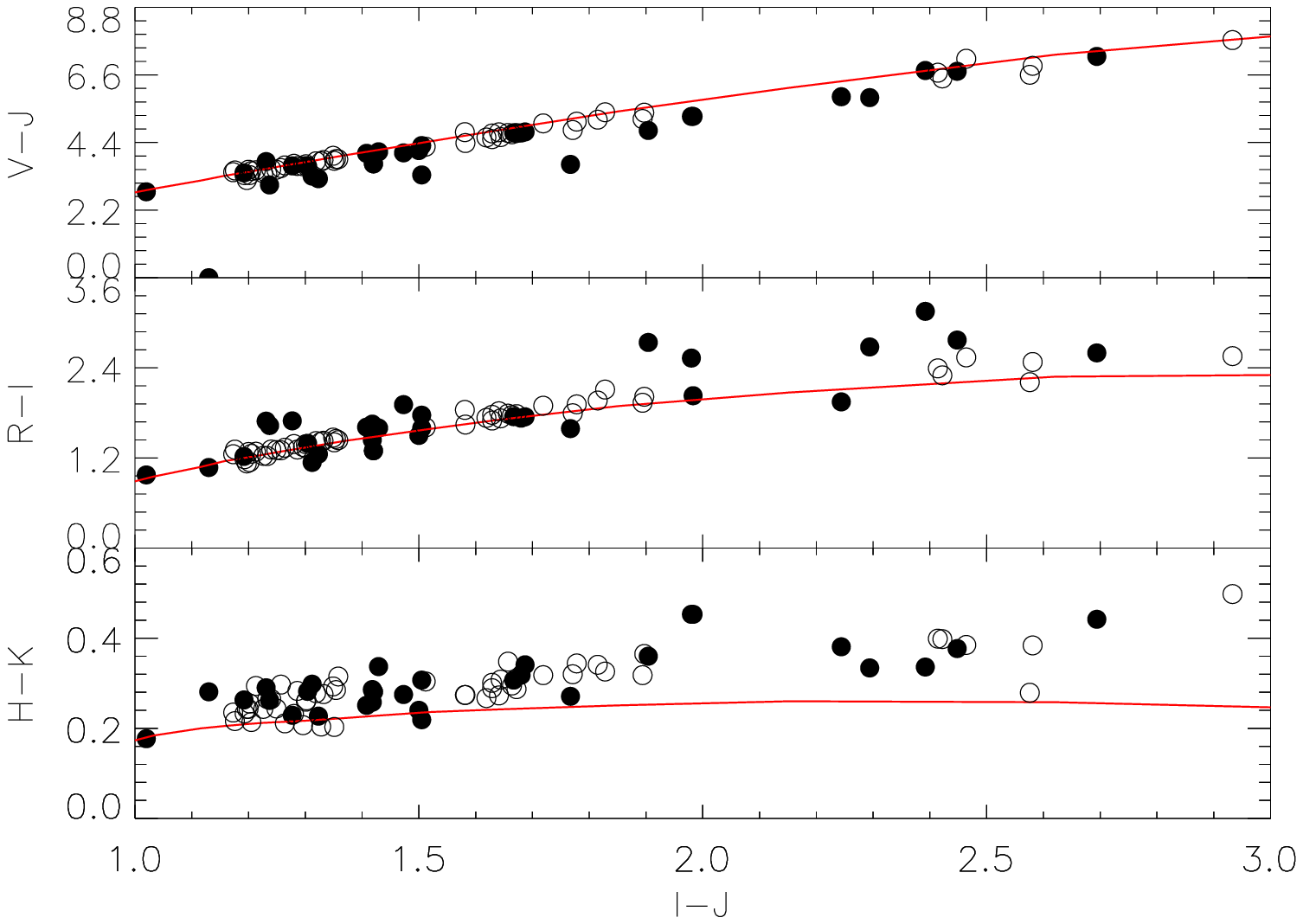}}
\caption{Optical and NIR colors obtained with the 5 Gyrs isochrones from \cite{BCAH98} at solar metallicity compared with the
   two observation samples (filled circles for the NTT sample and open circle for the SSO 2.3~m spectra).  Typical error bars are
   comparable or smaller than the size of the symbols.}
    \label{Fig:4}
 \end{figure*}

\begin{figure*}[ht!]
\ContinuedFloat
 \centering
   \subfloat{\label{fig:4}\includegraphics[width=12.5cm,height=10cm]{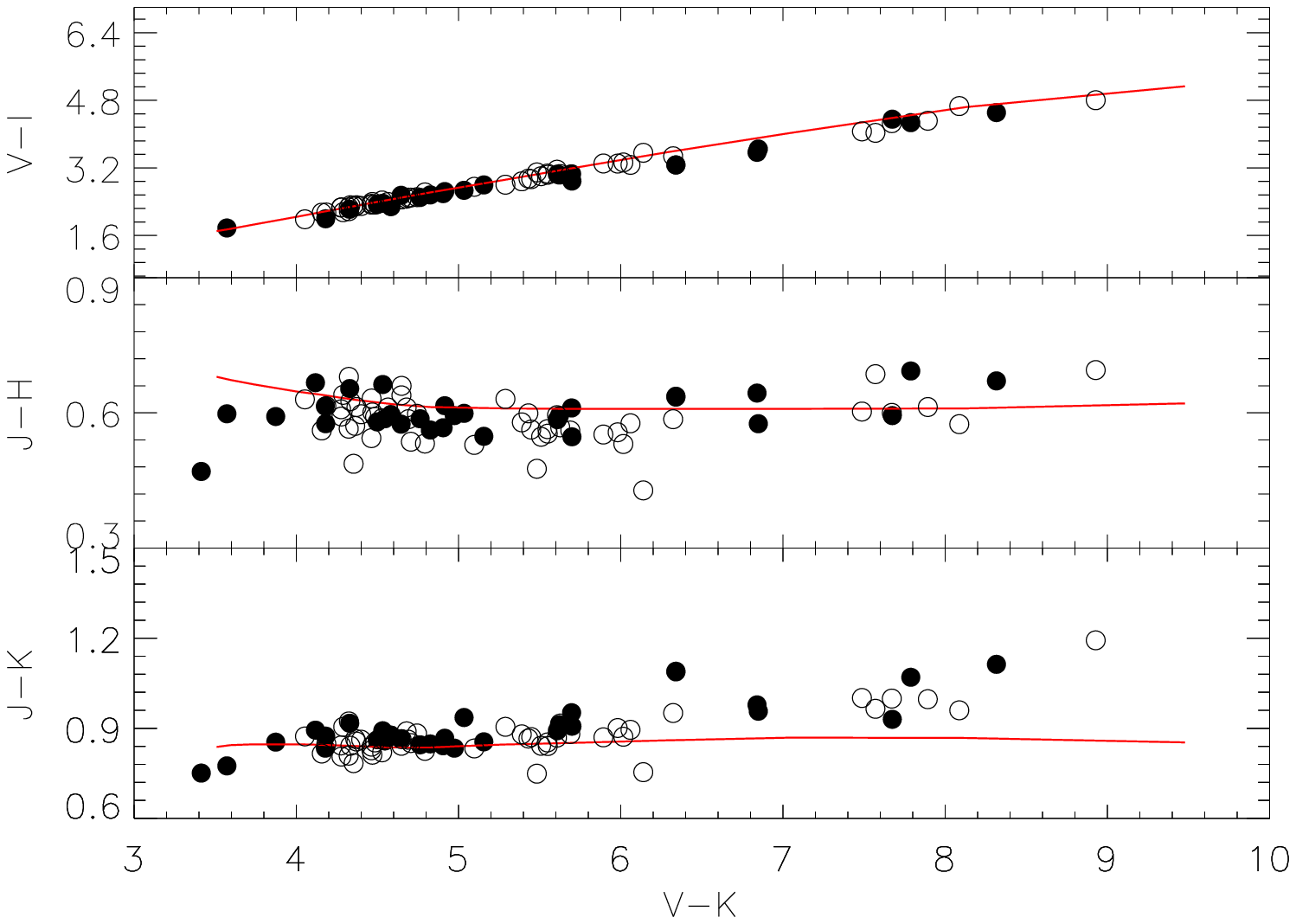}}
\qquad
   \subfloat{\label{fig:4}\includegraphics[width=12.5cm,height=10cm]{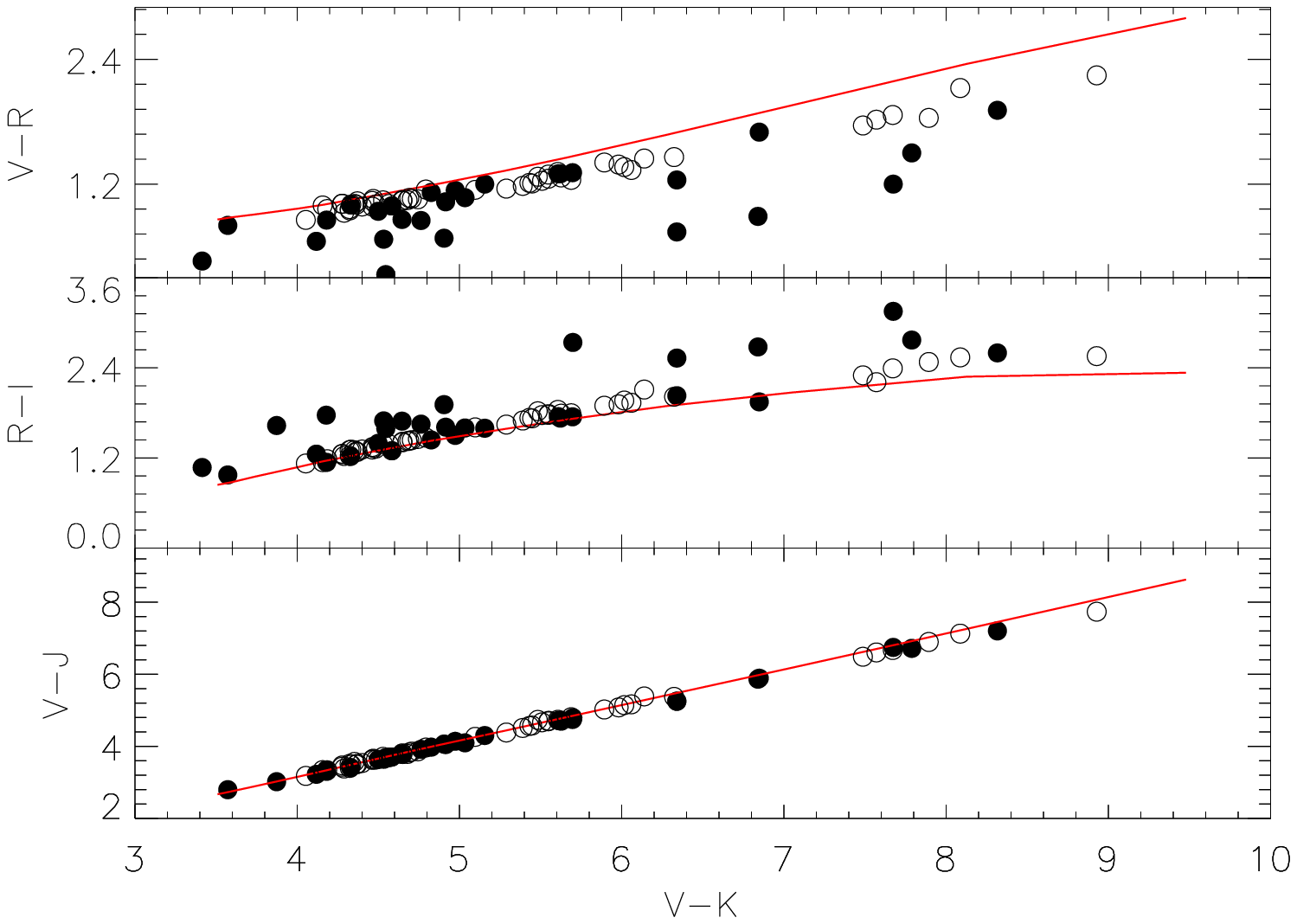}}
   \caption{Continued.}
     \label{Fig:4}
 \end{figure*}

\subsection{Photometric confrontation}

The models can be validated by comparing published isochrones interpolated into the new BT-Settl synthetic color tables with observed photometry.
We have taken the log g and $\teff$ for the fixed age of 5 Gyrs from \cite{BCAH98} isochrones and calculated the colors of the star
according to the BT-Settl models.
.
The models are compared to observations in color-color diagrams in Fig. ~\ref{Fig:4} for our two samples. The compiled photometry in the
NTT sample is less homogeneous, translating to a larger spread in particular for colors including the $V$ and $R$-band. This dispersion becomes dramatical for the coolest, and faintest, stars. 

except for lowest mass objects at very young ages. The isochrone reproduces the two samples over the entire M dwarf spectral range in most colors.
In particular, the models reproduce the $V$-band colors of M
dwarfs, as illustrated by the $V-I$, $V-J$ and $V-K$ colors. 
An increasing offset to the latest types persists in the $H-K$ and $V-R$ color indices. The observations  suggest also a flattening and possibly a rise in $J-H$ and $J-K$ to the latest types which is not reproduced by the model. These inadequacies at the coolest temperature could be linked to missing opacities.

\subsection{The $\teff$-scale of M dwarfs}
\label{S_teff}

\begin{figure*}[ht!s]
  \centering
  \includegraphics[width=15cm,height=10cm]{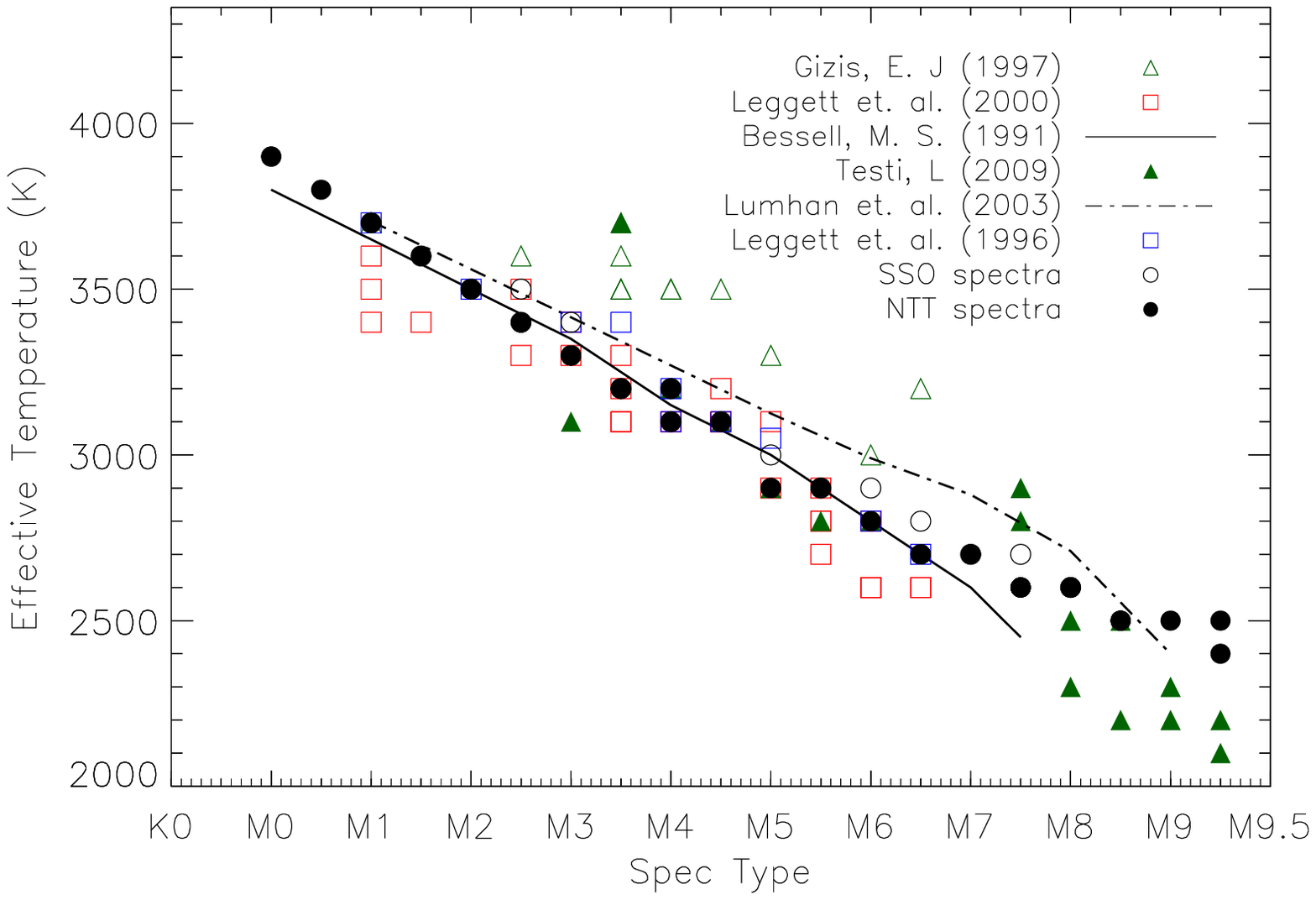}
  \caption{  
   Spectral type - $\teff$ relation obtained with the NTT sample (filled circles) and the SSO 2.3~m sample (open circles) compared
   to relations by \cite{Bessell1991}, \cite{Gizis1997}, \cite{Leggett1996}, \cite{Leggett2000}, \cite{Testi2009}, and \cite{Luhman1999}.
 }
\label{Fig:5}
   \end{figure*}

The effective temperature scale versus spectral type is shown in Fig.~\ref{Fig:5}. The $\teff$-scale determined using the NTT sample (filled
circles) is in agreement with the SSO sample (filled triangles) but we found systematically 100\,K higher $\teff$ for SSO sample for spectral
type later than M5. The relation shows a saturation trend for spectral types later than M8. This illustrate the fact that the optical spectrum no
longer change sensibly with $\teff$ in this regime due to dust formation. 

In the following we compare our scale to other works. \cite{Bessell1991} determined the temperatures by comparing blackbodies to the NIR photometry
of their sample. They used the temperature
calibration of \cite{Wing1979} and \cite{Veeder1974}. These calibrations were identical between  $ 2700 \le \teff \le 3500\,K$.
Their scale agrees with the modern values for M dwarfs earlier than M6, but becomes gradually too cool with later spectral type and too hot for
earlier M types. 

\cite{Leggett1996} used the Base grid by \cite{AH95} covering the range of parameter down to the coolest known M dwarfs, M subdwarfs and
brown dwarfs. They obtained the $\teff$ of M dwarfs by comparing the observed spectra to the synthetic spectra. They perform their comparison
independently at each of their four wavelength regions:  red, $J$, $H$, and $K$. The different wavelength regions gave consistent values of
$\teff$ within 300\,K. 
\cite{Gizis1997} used the NextGen model atmosphere grid by \cite{Allard1997}. These models include more molecular lines from ab
initio simulations (in particular for water vapor)  than the previous Base model grid. \cite{Leggett2000} used the
more modern AMES-Dusty model atmosphere grid by \cite{Allard2001}. They obtained a revised $\teff$ scale which is 150-200\,K cooler for
early-Ms, and 200\,K hotter for late-Ms than the scale presented in Fig.~\ref{Fig:5}. \cite{Testi2009} determine the $\teff$ by fitting the
synthetic spectra to the observations. They used three classes of models: the AMES-Dusty, AMES-Cond and the BT-Settl models. With some individual
exceptions they found that the BT-Settl models were the most appropriate for M type and early L-type dwarfs.

Finally, for spectral type later than M0, \cite{Luhman2003} adopted the effective temperature which is based on the NextGen and AMES-Dusty
evolutionary models of \cite{BCAH98} and \cite{Chabrier2000} respectively. They obtained the $\teff$ by comparing the H-R diagram from
theoretical isochrones of \cite{BCAH98} and \cite{Chabrier2000}. For M8 and M9, \cite{Luhman2003} adjusted the temperature scale from
\cite{Luhman1999} so that spectral sequence fall parallel to the isochrones. Their $\teff$ conversion is likely to be inaccurate at some level,
but as it falls between the scales for dwarfs and giants, the error in $\teff$ are modest.

The different $\teff$ scales are in agreement within 250-300\,K. But the \cite{Gizis1997} relation shows the largest differences, with the largest
$\teff$-values (up to 500\,K). This is due to the incompleteness of the TiO and water vapor line lists used in the NextGen model atmospheres. Note
also how the \cite{Luhman2003} $\teff$ scale is gradually overestimating $\teff$ towards the bottom of the main sequence for spectral types later
than M4.

\medskip

\begin{figure*}[ht!]
   \centering
   \subfloat{\label{fig:6}\includegraphics[width=12.8cm,height=7.5cm]{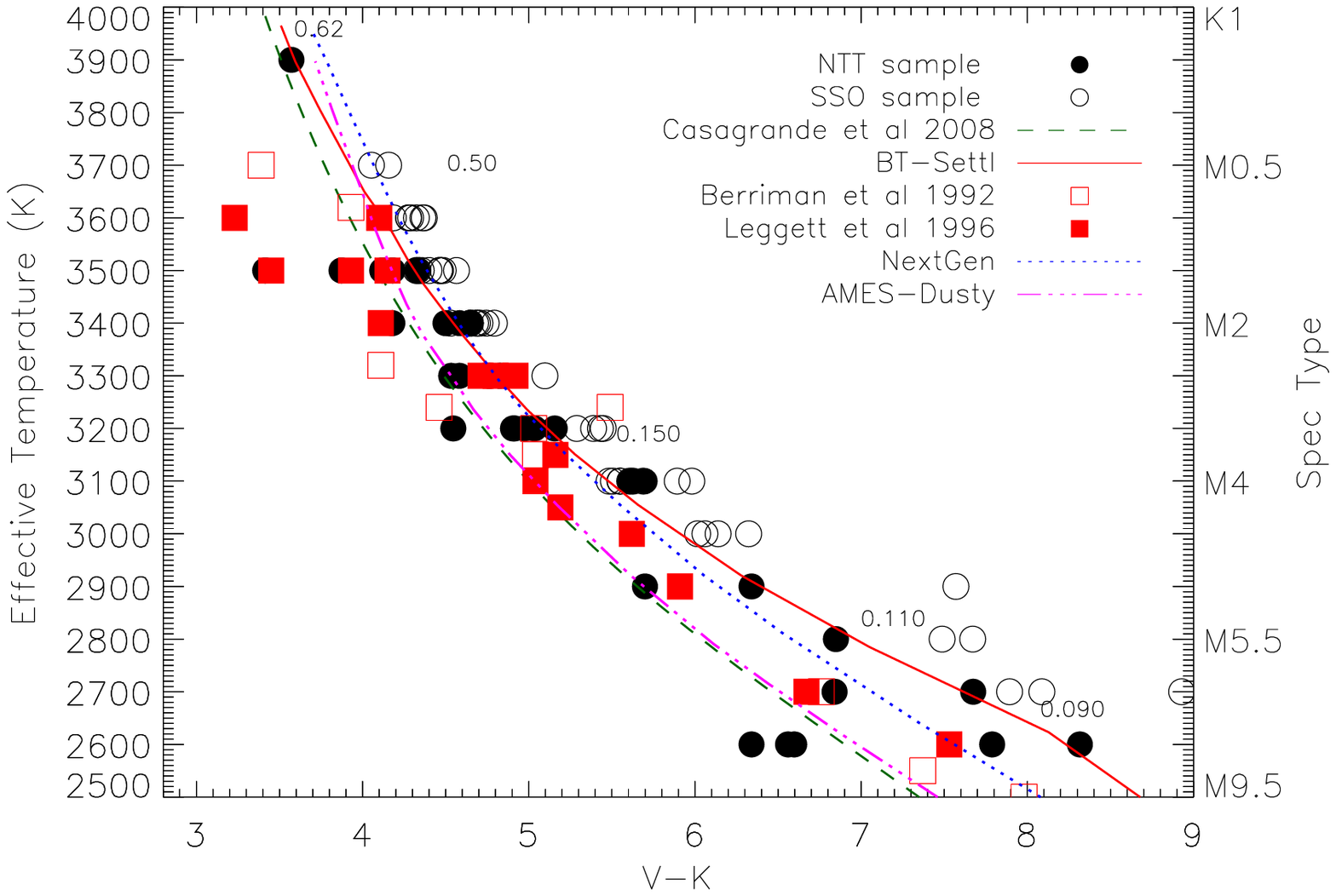}}
\qquad
   \subfloat{\label{fig:6}\includegraphics[width=12.8cm,height=7.5cm]{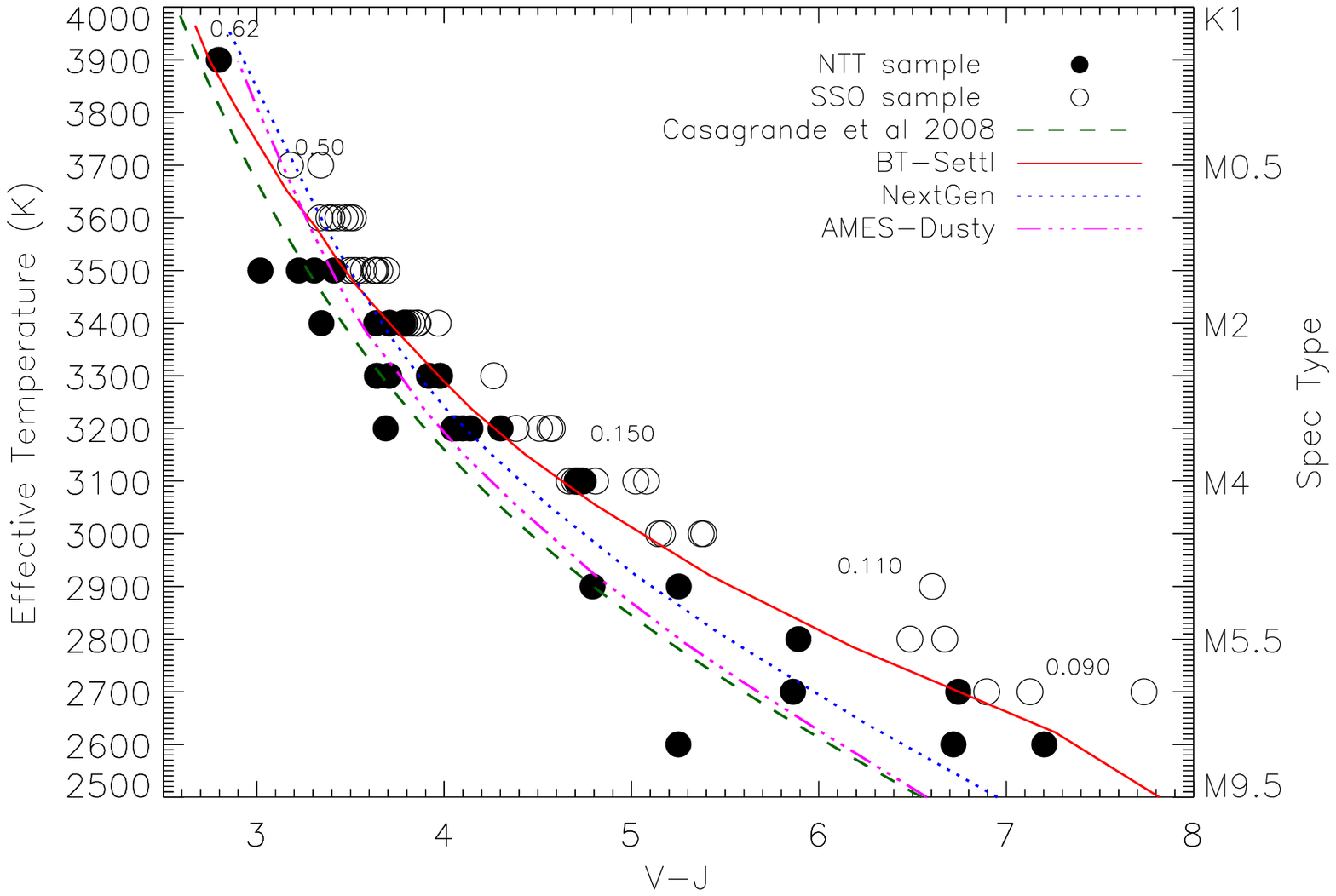}}
\qquad
   \subfloat{\label{fig:6}\includegraphics[width=12.8cm,height=7.5cm]{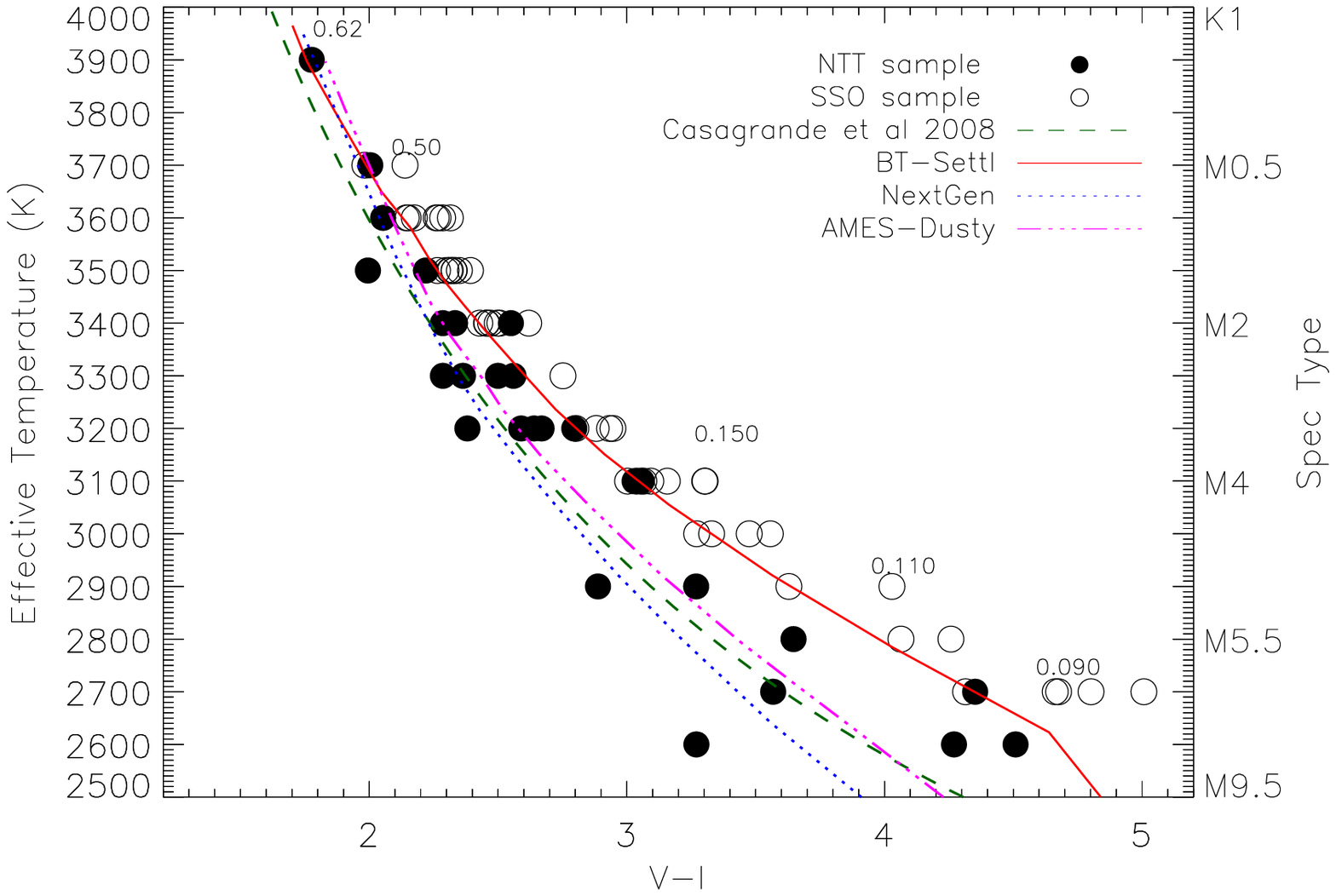}}
\caption{Color-$\teff$ plots in different bands from the NTT sample (filled circles) and the SSO 2.3~m sample (open circles).
Spectral types are also indicated. The predictions from BT-Settl (solid line), NextGen (dotted line) and AMES-Dusty (das-dotted) for
solar metallicities are over plotted. Theoretical masses in solar mass are indicated. Predictions from other authors are shown for comparison when
available.}
\label{Fig:6}
   \end{figure*}
\begin{figure*}[ht!]
\ContinuedFloat
   \centering
   \subfloat{\label{fig:6}\includegraphics[width=12.5cm,height=7.5cm]{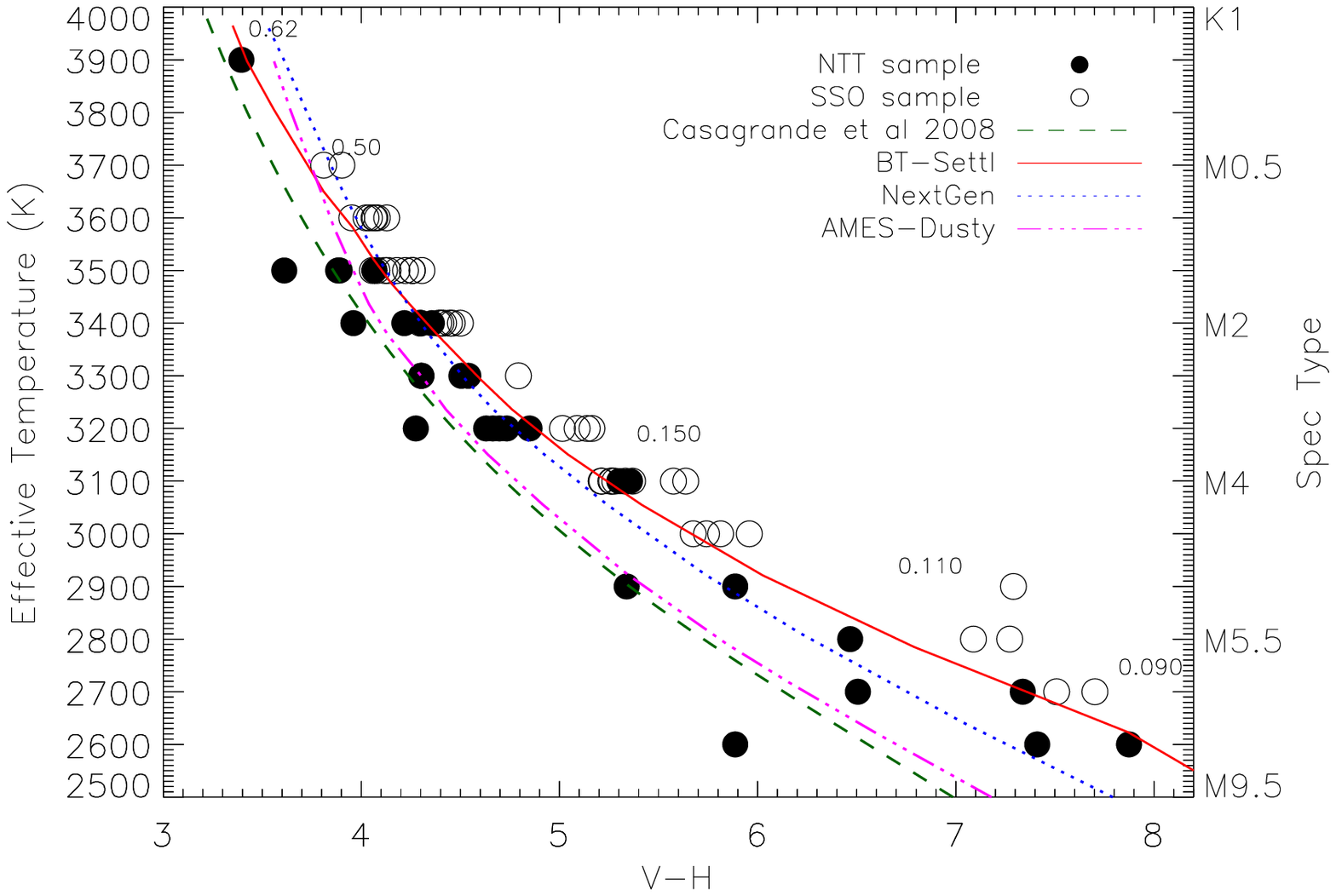}}
\qquad
   \subfloat{\label{fig:6}\includegraphics[width=12.5cm,height=7.5cm]{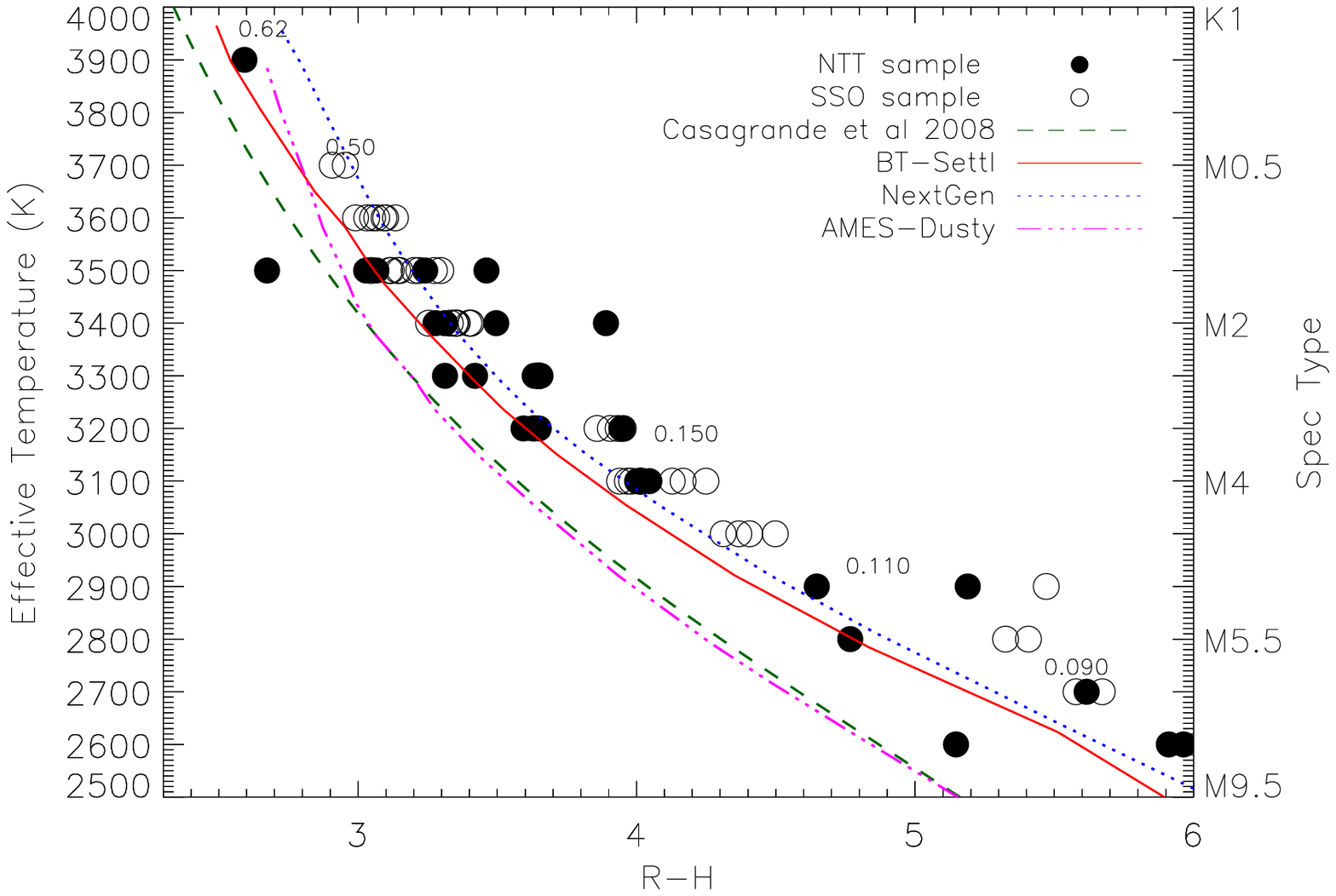}}
\qquad
   \subfloat{\label{fig:6}\includegraphics[width=12.5cm,height=7.5cm]{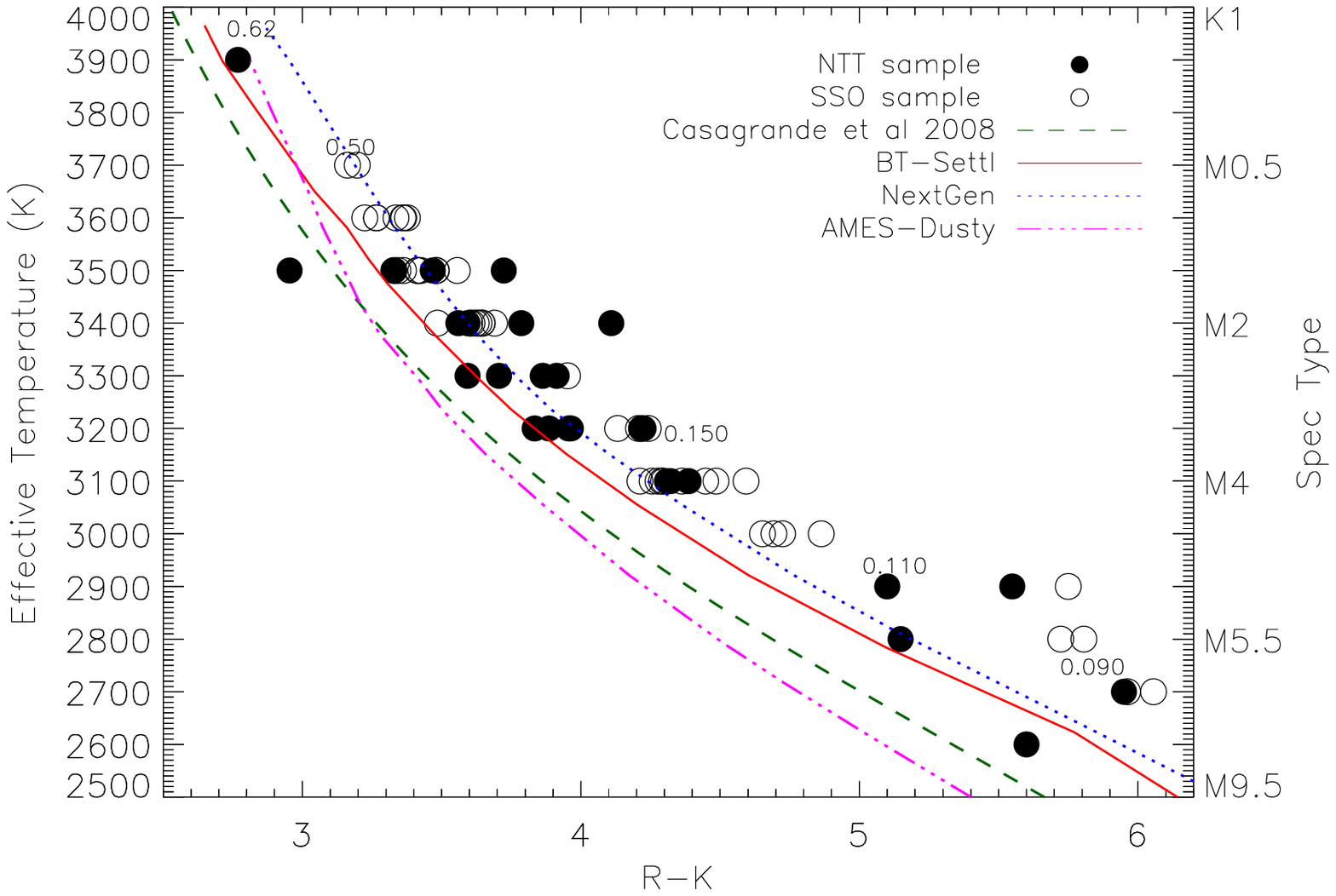}}
\caption{Continued.}
\label{Fig:6}
   \end{figure*}
\begin{figure*}[ht!]
\ContinuedFloat
   \centering
   \subfloat{\label{fig:6}\includegraphics[width=12.5cm,height=7.5cm]{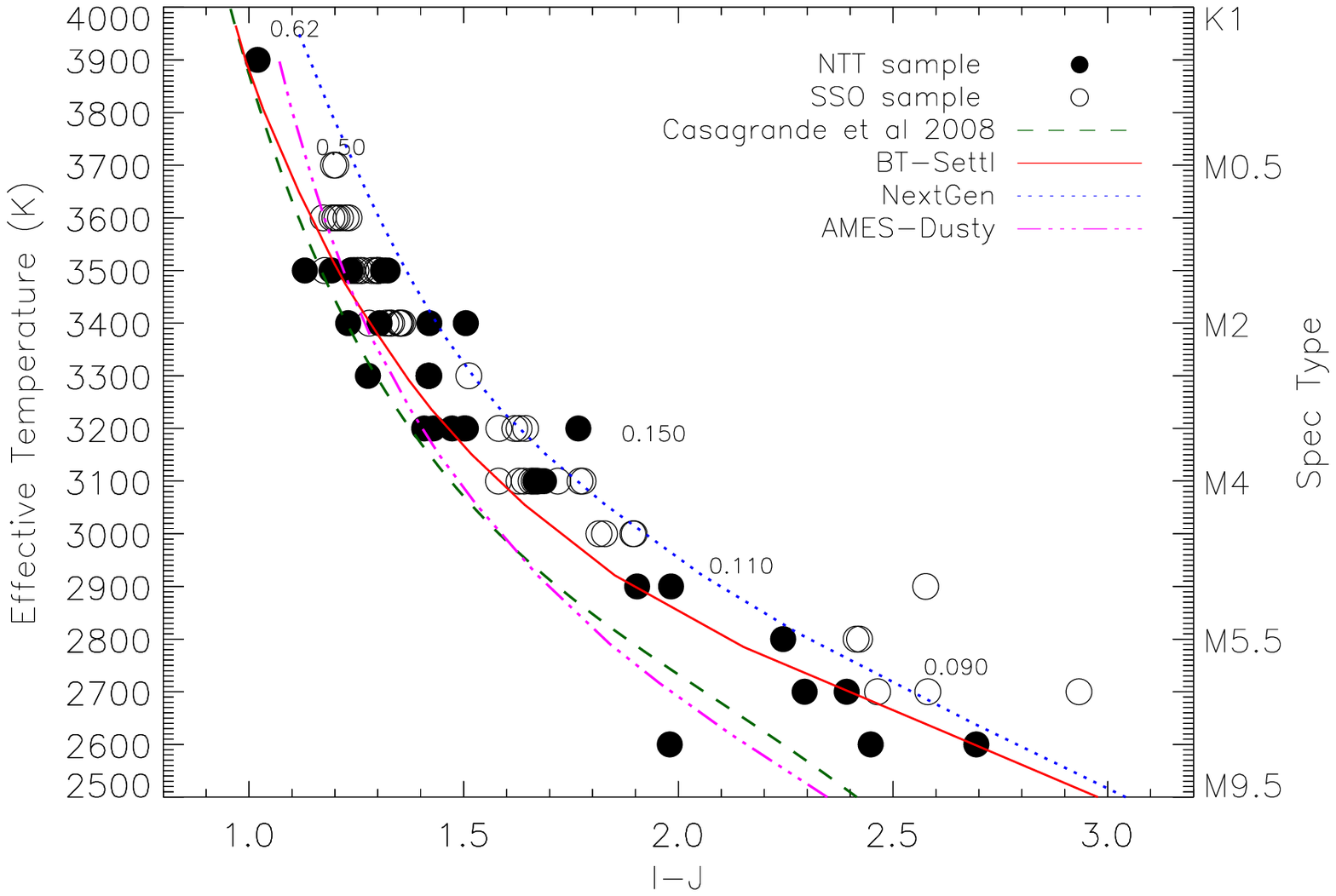}}
\qquad
   \subfloat{\label{fig:6}\includegraphics[width=12.5cm,height=7.5cm]{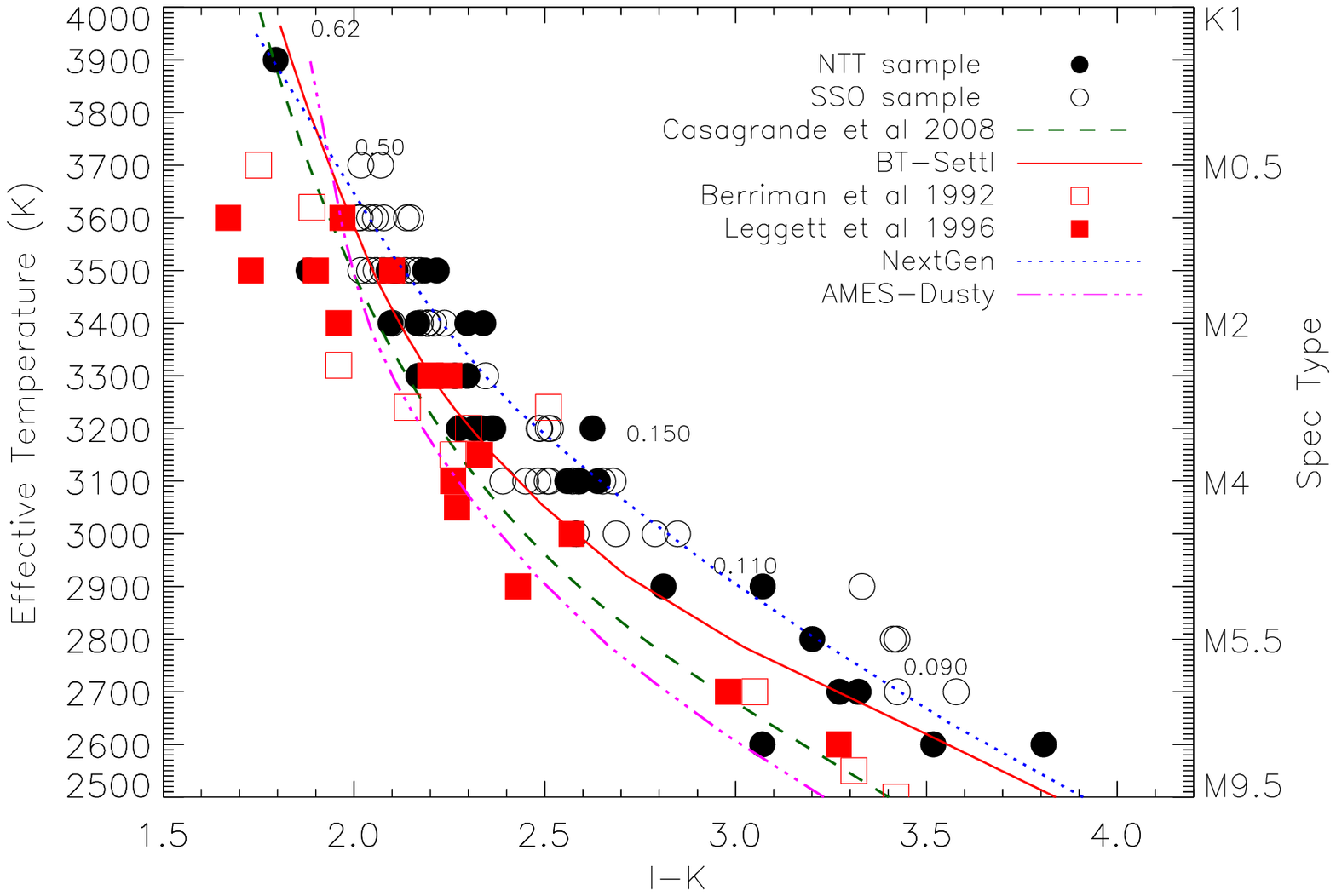}}
\qquad
   \subfloat{\label{fig:6}\includegraphics[width=12.5cm,height=7.5cm]{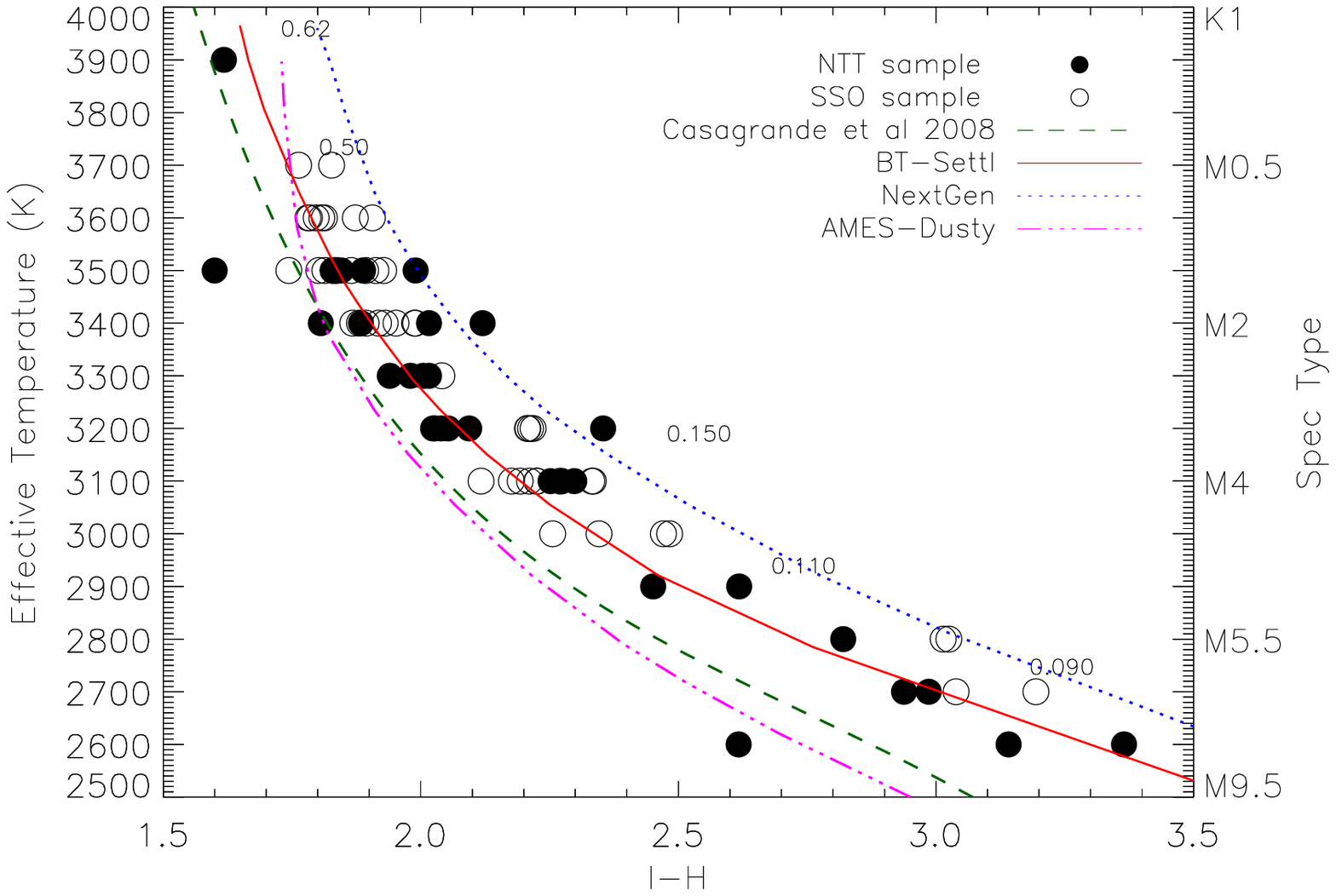}}
   \caption{Continued.}
   \label{Fig:6}
   \end{figure*}

$\teff$ versus color relations are shown in Fig.~\ref{Fig:6} in various photometric bands. The photometry of our NTT sample (filled circles) is
compiled from the literature, causing a large spread particularly in the $V$ and $R$-band. The SSO 2.3~m sample (filled triangles) in
comparison is more uniform. Our relations are compared to the predictions from BT-Settl isochrones at 5 Gyrs. It shows that the model is able to
reproduce quite
properly the colors of M-dwarfs, even in the $V$-band. There is a slight offset visible in the R-band due to missing molecular opacities (see above).
These relations are compared to previously published relations when available.

\cite{Berriman1992} derive the $\teff$ by matching the blackbody flux anchored at $K$ band (2.2~$\mu$m) to the total bolometric flux including
both the spectroscopic and photometric observed data points. They estimated the uncertainties in $\teff$ to be $\pm$ 4$\%$. \cite{Leggett1996} used
the synthetic $I-K$ and $I-J$ colors to estimate $\teff$.
\cite{Leggett1996} used synthetic broadband colors from the preliminary version of AMES-Dusty model produced by \cite{Allard1994}. They used the $V-K$, $I-K$, $J-H$ and $H-K$ colors assuming log$g = 5.0$ and solar metallicity, and found a hotter $\teff$-scale (by on average of 130\,K) than that of \cite{Berriman1992}.  
More recently, \cite{Casagrande2008} used the \texttt{PHOENIX} Cond-GAIA model atmosphere grid (P. H. Hauschildt, unpublished) to determine the atmospheric parameters of their sample of 343 nearby M dwarfs with high-quality optical and IR photometry. These models are similar to those
published by \cite{Allard2001} with the exception that they were computed by solving the radiative transfer in spherical symmetry. The authors
determined the $\teff$ using a version of the multiple optical-infrared method (IRFM) generalized to M dwarfs, and elaborated by \cite{Blackwell1977}
and \cite{Blackwell1979,Blackwell1980}. Fig.~\ref{Fig:6} shows that the \cite{Casagrande2008} $\teff$-scale is systematically, and progressively with
decreasing $\teff$, cooler than the BT-Settl isochrones. Given that a large number of stars are common with \cite{Casagrande2008} sample, we did a
star-by-star comparison of the $\teff$ determination. The values are given in Table~1 and 2. It confirms the systematic offset in the temperature
scale.
For cooler stars with $\teff < 3000$\,K, the $\teff$ determinations diverge by 100 to 300\,K. 

This is due, among other things, to the use of the \cite{GNS93} solar elemental abundances (see Allard et al. 2012\nocite{Allard2012} for a comparison
of the different solar elemental abundance determinations and their effects on model atmospheres).

\section{Conclusion}
\label{S_concl}

We have compared a revised version of the BT-Settl model atmospheres \citep{Allard2012} to the observed NTT and SSO 2.3~m spectra and colors. 
This new version uses the \cite{Caffau2011} solar elemental abundances, updates to the atomic and molecular line broadening and the TiO line
list from \cite[][and B. Plez, private communication]{Plez1998}. This list provides a more accurate description on the TiO bands in the M dwarfs. The
systematic discrepancy
 between the delta- and epsilon-bands found by \cite{Reiners2005}, which seriously affected the effective temperature determination, is largely
alleviated by using the \cite[][and B. Plez, private communication]{Plez1998} TiO line list although discrepancies remain for the coolest stars.
The BT-Settl models reproduce the spectral energy distribution and observed colors across the M dwarfs spectral regime to an unprecedented
quality, as well as the colors. The $V$ band is also well reproduced by the models. Some
discrepancies remain in the strength of some and missing other molecular absorption bands in particular in the ultraviolet spectral range. 

Effective temperatures were determined by using a least-square minimization routine which gives accurate temperatures within 100\,K uncertainty. We
compare our temperature versus color relations using multi-wavelength photometry with  the predictions from BT-Settl isochrones, assuming an age of 5
Gyrs. In general, the
BT-Settl isochrones are in good agreement with the observed colors, even at temperatures below  2800\,K  affected by dust-treatment in the BT-Settl
models. We found that the \cite{Casagrande2008} $\teff$-scale is  systematically  cooler than the BT-Settl isochrones due, among other things,
to  the \cite{GNS93} solar elemental abundances adopted in the GAIA-Cond model atmosphere grid used for that work.
The \cite{Luhman2003} $\teff$-scale is on the contrary progressively too hot towards the bottom of the main sequence.  New interior and evolution
models are currently being prepared, based on the BT-Settl models.

We provide and compare temperature versus color relations in the Optical and Infrared which matches well the BT-Settl isochrones and can be
further used for large photometric datasets. We determined the effective temperature scale for the M dwarfs in our samples.  Our effective
temperature scale extended down to the latest-type M dwarfs where the dust cloud begins to form in their atmosphere.

\begin{acknowledgements}
The use of Simbad and Vizier databases at CDS, as well as the 
ARICNS database was very helpful for this research.
The research leading to these results has received funding from the 
French ``Agence Nationale de la Recherche'' (ANR), the ``Programme 
National de Physique Stellaire'' (PNPS) of CNRS (INSU), and the
European Research Council under the European Community's Seventh 
Framework Programme (FP7/2007-2013 Grant Agreement no. 247060).
It was also conducted within the Lyon Institute of Origins under 
grant ANR-10-LABX-66. 

The computations were performed at
the {\sl P\^ole Scientifique de Mod\'elisation Num\'erique} (PSMN) at
the {\sl \'Ecole Normale Sup\'erieure} (ENS) in Lyon, and at the {\sl
Gesellschaft f{\"u}r Wissenschaftliche Datenverarbeitung
G{\"o}ttingen} in collaboration with the Institut f{\"u}r Astrophysik
G{\"o}ttingen.

\end{acknowledgements}

\bibliographystyle{aa}
\bibliography{ref}

\begin{thebibliography}{91}
\expandafter\ifx\csname natexlab\endcsname\relax\def\natexlab#1{#1}\fi

\bibitem[{{Abel} {et~al.}(2011){Abel}, {Frommhold}, {Li}, \& {Hunt}}]{Abel2011}
{Abel}, M., {Frommhold}, L., {Li}, X., \& {Hunt}, K.~L.~C. 2011, in 66th
  International Symposium On Molecular Spectroscopy

\bibitem[{{Allard}(1990)}]{AllardPhDT90}
{Allard}, F. 1990, PhD thesis, Ruprecht Karls Univ.~Heidelberg

\bibitem[{{Allard} \& {Hauschildt}(1995)}]{AH95}
{Allard}, F. \& {Hauschildt}, P.~H. 1995, \apj, 445, 433

\bibitem[{{Allard} {et~al.}(1997){Allard}, {Hauschildt}, {Alexander}, \&
  {Starrfield}}]{Allard1997}
{Allard}, F., {Hauschildt}, P.~H., {Alexander}, D.~R., \& {Starrfield}, S.
  1997, \araa, 35, 137

\bibitem[{{Allard} {et~al.}(2001){Allard}, {Hauschildt}, {Alexander},
  {Tamanai}, \& {Schweitzer}}]{Allard2001}
{Allard}, F., {Hauschildt}, P.~H., {Alexander}, D.~R., {Tamanai}, A., \&
  {Schweitzer}, A. 2001, \apj, 556, 357

\bibitem[{{Allard} {et~al.}(1994){Allard}, {Hauschildt}, {Miller}, \&
  {Tennyson}}]{Allard1994}
{Allard}, F., {Hauschildt}, P.~H., {Miller}, S., \& {Tennyson}, J. 1994, \apjl,
  426, L39

\bibitem[{{Allard} \& {Homeier}(2012)}]{Allard2012b}
{Allard}, F. \& {Homeier}, D. 2012, ArXiv e-prints

\bibitem[{{Allard} {et~al.}(2011){Allard}, {Homeier}, \&
  {Freytag}}]{Allard2011}
{Allard}, F., {Homeier}, D., \& {Freytag}, B. 2011, in Astronomical Society of
  the Pacific Conference Series, Vol. 448, Astronomical Society of the Pacific
  Conference Series, ed. C.~{Johns-Krull}, M.~K. {Browning}, \& A.~A. {West},
  91

\bibitem[{{Allard} {et~al.}(2012{\natexlab{a}}){Allard}, {Homeier}, \&
  {Freytag}}]{Allard2012}
{Allard}, F., {Homeier}, D., \& {Freytag}, B. 2012{\natexlab{a}}, Royal Society
  of London Philosophical Transactions Series A, 370, 2765

\bibitem[{{Allard} {et~al.}(2012{\natexlab{b}}){Allard}, {Homeier}, {Freytag},
  \& {Sharp}}]{Allard2012c}
{Allard}, F., {Homeier}, D., {Freytag}, B., \& {Sharp}, C.~M.
  2012{\natexlab{b}}, in EAS Publications Series, Vol.~57, EAS Publications
  Series, ed. C.~{Reyl{\'e}}, C.~{Charbonnel}, \& M.~{Schultheis}, 3--43

\bibitem[{{Allard} {et~al.}(2007){Allard}, {Kielkopf}, \&
  {Allard}}]{AllardN2007}
{Allard}, N.~F., {Kielkopf}, J.~F., \& {Allard}, F. 2007, European Physical
  Journal D, 44, 507

\bibitem[{{Asplund} {et~al.}(2009){Asplund}, {Grevesse}, {Sauval}, \&
  {Scott}}]{Asplund2009}
{Asplund}, M., {Grevesse}, N., {Sauval}, A.~J., \& {Scott}, P. 2009, \araa, 47,
  481

\bibitem[{{Bailey} \& {Kedziora-Chudczer}(2012)}]{VSTAR2012}
{Bailey}, J. \& {Kedziora-Chudczer}, L. 2012, \mnras, 419, 1913

\bibitem[{{Baraffe} {et~al.}(1998){Baraffe}, {Chabrier}, {Allard}, \&
  {Hauschildt}}]{BCAH98}
{Baraffe}, I., {Chabrier}, G., {Allard}, F., \& {Hauschildt}, P.~H. 1998, \aap,
  337, 403

\bibitem[{{Barber} {et~al.}(2006){Barber}, {Tennyson}, {Harris}, \&
  {Tolchenov}}]{Barber2006}
{Barber}, R.~J., {Tennyson}, J., {Harris}, G.~J., \& {Tolchenov}, R.~N. 2006,
  \mnras, 368, 1087

\bibitem[{{Barklem} {et~al.}(2000){Barklem}, {Piskunov}, \&
  {O'Mara}}]{Barklem2000}
{Barklem}, P.~S., {Piskunov}, N., \& {O'Mara}, B.~J. 2000, \aap, 363, 1091

\bibitem[{{Berriman} \& {Reid}(1987)}]{Berriman1987}
{Berriman}, G. \& {Reid}, N. 1987, \mnras, 227, 315

\bibitem[{{Berriman} {et~al.}(1992){Berriman}, {Reid}, \&
  {Leggett}}]{Berriman1992}
{Berriman}, G., {Reid}, N., \& {Leggett}, S.~K. 1992, \apjl, 392, L31

\bibitem[{{Bessell}(1991)}]{Bessell1991}
{Bessell}, M.~S. 1991, \aj, 101, 662

\bibitem[{{Bessell}(1995)}]{Bessell1995}
{Bessell}, M.~S. 1995, in The Bottom of the Main Sequence - and Beyond, ed.
  C.~G. {Tinney}, 123

\bibitem[{{Blackwell} {et~al.}(1980){Blackwell}, {Petford}, \&
  {Shallis}}]{Blackwell1980}
{Blackwell}, D.~E., {Petford}, A.~D., \& {Shallis}, M.~J. 1980, \aap, 82, 249

\bibitem[{{Blackwell} \& {Shallis}(1977)}]{Blackwell1977}
{Blackwell}, D.~E. \& {Shallis}, M.~J. 1977, \mnras, 180, 177

\bibitem[{{Blackwell} {et~al.}(1979){Blackwell}, {Shallis}, \&
  {Selby}}]{Blackwell1979}
{Blackwell}, D.~E., {Shallis}, M.~J., \& {Selby}, M.~J. 1979, \mnras, 188, 847

\bibitem[{{Bochanski} {et~al.}(2010){Bochanski}, {Hawley}, {Covey}, {West},
  {Reid}, {Golimowski}, \& {Ivezi{\'c}}}]{Bochanski2010}
{Bochanski}, J.~J., {Hawley}, S.~L., {Covey}, K.~R., {et~al.} 2010, \aj, 139,
  2679

\bibitem[{{Bonfils} {et~al.}(2012){Bonfils}, {Gillon}, {Udry}, {Armstrong},
  {Bouchy}, {Delfosse}, {Forveille}, {Jehin}, {Lendl}, {Lovis}, {Mayor},
  {McCormac}, {Neves}, {Pepe}, {Perrier}, {Pollaco}, {Queloz}, \&
  {Santos}}]{Bonfils2012}
{Bonfils}, X., {Gillon}, M., {Udry}, S., {et~al.} 2012, ArXiv e-prints

\bibitem[{{Bonfils} {et~al.}(2007){Bonfils}, {Mayor}, {Delfosse}, {Forveille},
  {Gillon}, {Perrier}, {Udry}, {Bouchy}, {Lovis}, {Pepe}, {Queloz}, {Santos},
  \& {Bertaux}}]{Bonfils2007}
{Bonfils}, X., {Mayor}, M., {Delfosse}, X., {et~al.} 2007, \aap, 474, 293

\bibitem[{Borysow {et~al.}(2001)Borysow, J{\o}rgensen, \& Fu}]{Borysow2001}
Borysow, A., J{\o}rgensen, U.~G., \& Fu, Y. 2001, Journal of Quantitative
  Spectroscopy and Radiative Transfer, 68, 235

\bibitem[{{Caffau} {et~al.}(2011){Caffau}, {Ludwig}, {Steffen}, {Freytag}, \&
  {Bonifacio}}]{Caffau2011}
{Caffau}, E., {Ludwig}, H.-G., {Steffen}, M., {Freytag}, B., \& {Bonifacio}, P.
  2011, \solphys, 268, 255

\bibitem[{{Casagrande} {et~al.}(2008){Casagrande}, {Flynn}, \&
  {Bessell}}]{Casagrande2008}
{Casagrande}, L., {Flynn}, C., \& {Bessell}, M. 2008, \mnras, 389, 585

\bibitem[{{Casagrande} \& {Sch{\"o}nrich}(2012)}]{Casagrande2012}
{Casagrande}, L. \& {Sch{\"o}nrich}, R. 2012, in European Physical Journal Web
  of Conferences, Vol.~19, European Physical Journal Web of Conferences, 5004

\bibitem[{{Chabrier}(2003)}]{Chabrier03}
{Chabrier}, G. 2003, \pasp, 115, 763

\bibitem[{{Chabrier}(2005)}]{Chabrier05}
{Chabrier}, G. 2005, in Astrophysics and Space Science Library, Vol. 327, The
  Initial Mass Function 50 Years Later, ed. E.~{Corbelli}, F.~{Palla}, \&
  H.~{Zinnecker}, 41

\bibitem[{{Chabrier} {et~al.}(2000){Chabrier}, {Baraffe}, {Allard}, \&
  {Hauschildt}}]{Chabrier2000}
{Chabrier}, G., {Baraffe}, I., {Allard}, F., \& {Hauschildt}, P. 2000, \apj,
  542, 464

\bibitem[{{Chowdhury} {et~al.}(2006){Chowdhury}, {Merer}, {Rixon}, {Bernath},
  \& {Ram}}]{Chowdhury2006}
{Chowdhury}, P.~K., {Merer}, A.~J., {Rixon}, S.~J., {Bernath}, P.~F., \& {Ram},
  R.~S. 2006, Physical Chemistry Chemical Physics (Incorporating Faraday
  Transactions), 8, 822

\bibitem[{{Crifo} {et~al.}(2005){Crifo}, {Phan-Bao}, {Delfosse}, {Forveille},
  {Guibert}, {Mart{\'{\i}}n}, \& {Reyl{\'e}}}]{Crifo2005}
{Crifo}, F., {Phan-Bao}, N., {Delfosse}, X., {et~al.} 2005, \aap, 441, 653

\bibitem[{{Delfosse} {et~al.}(1999){Delfosse}, {Tinney}, {Forveille},
  {Epchtein}, {Borsenberger}, {Fouqu{\'e}}, {Kimeswenger}, \&
  {Tiph{\`e}ne}}]{Delfosse1999}
{Delfosse}, X., {Tinney}, C.~G., {Forveille}, T., {et~al.} 1999, \aaps, 135, 41

\bibitem[{{Dulick} {et~al.}(2003){Dulick}, {Bauschlicher}, {Burrows}, {Sharp},
  {Ram}, \& {Bernath}}]{Dulick2003}
{Dulick}, M., {Bauschlicher}, Jr., C.~W., {Burrows}, A., {et~al.} 2003, \apj,
  594, 651

\bibitem[{{Epchtein}(1997)}]{epchtein1997}
{Epchtein}, N. 1997, in Astrophysics and Space Science Library, Vol. 210, The
  Impact of Large Scale Near-IR Sky Surveys, ed. F.~{Garzon}, N.~{Epchtein},
  A.~{Omont}, B.~{Burton}, \& P.~{Persi}, 15

\bibitem[{{Freytag} {et~al.}(2012){Freytag}, {Steffen}, {Ludwig},
  {Wedemeyer-B{\"o}hm}, {Schaffenberger}, \& {Steiner}}]{Freytag2012}
{Freytag}, B., {Steffen}, M., {Ludwig}, H.-G., {et~al.} 2012, J. Comp. Phys.,
  231, 919

\bibitem[{{Gamache} {et~al.}(1996){Gamache}, {Lynch}, \& {Brown}}]{Gamache1996}
{Gamache}, R.~R., {Lynch}, R., \& {Brown}, L.~R. 1996, \jqsrt, 56, 471

\bibitem[{{Gizis}(1996)}]{Gizis1996}
{Gizis}, J.~E. 1996, in Astronomical Society of the Pacific Conference Series,
  Vol. 109, Cool Stars, Stellar Systems, and the Sun, ed. {R.~Pallavicini \&
  A.~K.~Dupree}, 683

\bibitem[{{Gizis}(1997)}]{Gizis1997}
{Gizis}, J.~E. 1997, \aj, 113, 806

\bibitem[{{Goorvitch} \& {Chackerian}(1994{\natexlab{a}})}]{Goorvitch94a}
{Goorvitch}, D. \& {Chackerian}, Jr., C. 1994{\natexlab{a}}, \apjs, 91, 483

\bibitem[{{Goorvitch} \& {Chackerian}(1994{\natexlab{b}})}]{Goorvitch94b}
{Goorvitch}, D. \& {Chackerian}, Jr., C. 1994{\natexlab{b}}, \apjs, 92, 311

\bibitem[{{Grevesse} {et~al.}(1993){Grevesse}, {Noels}, \& {Sauval}}]{GNS93}
{Grevesse}, N., {Noels}, A., \& {Sauval}, A.~J. 1993, \aap, 271, 587

\bibitem[{{Hauschildt} {et~al.}(1999){Hauschildt}, {Allard}, \&
  {Baron}}]{Hauschildt1999}
{Hauschildt}, P.~H., {Allard}, F., \& {Baron}, E. 1999, \apj, 512, 377

\bibitem[{{Hauschildt} {et~al.}(1997){Hauschildt}, {Baron}, \&
  {Allard}}]{Phoenix97}
{Hauschildt}, P.~H., {Baron}, E., \& {Allard}, F. 1997, \apj, 483, 390

\bibitem[{{Kirkpatrick} {et~al.}(1993){Kirkpatrick}, {Kelly}, {Rieke},
  {Liebert}, {Allard}, \& {Wehrse}}]{Kirkpatrick1993}
{Kirkpatrick}, J.~D., {Kelly}, D.~M., {Rieke}, G.~H., {et~al.} 1993, \apj, 402,
  643

\bibitem[{{Koen} \& {Eyer}(2002)}]{Koen2002}
{Koen}, C. \& {Eyer}, L. 2002, \mnras, 331, 45

\bibitem[{{Koen} {et~al.}(2010){Koen}, {Kilkenny}, {van Wyk}, \&
  {Marang}}]{Koen2010}
{Koen}, C., {Kilkenny}, D., {van Wyk}, F., \& {Marang}, F. 2010, \mnras, 403,
  1949

\bibitem[{{Leggett} {et~al.}(1996){Leggett}, {Allard}, {Berriman}, {Dahn}, \&
  {Hauschildt}}]{Leggett1996}
{Leggett}, S.~K., {Allard}, F., {Berriman}, G., {Dahn}, C.~C., \& {Hauschildt},
  P.~H. 1996, \apjs, 104, 117

\bibitem[{{Leggett} {et~al.}(2000){Leggett}, {Allard}, {Dahn}, {Hauschildt},
  {Kerr}, \& {Rayner}}]{Leggett2000}
{Leggett}, S.~K., {Allard}, F., {Dahn}, C., {et~al.} 2000, \apj, 535, 965

\bibitem[{{Leggett} {et~al.}(1998){Leggett}, {Allard}, \&
  {Hauschildt}}]{Leggett1998}
{Leggett}, S.~K., {Allard}, F., \& {Hauschildt}, P.~H. 1998, \apj, 509, 836

\bibitem[{{Ludwig} {et~al.}(2002){Ludwig}, {Allard}, \&
  {Hauschildt}}]{Ludwig2002}
{Ludwig}, H.-G., {Allard}, F., \& {Hauschildt}, P.~H. 2002, \aap, 395, 99

\bibitem[{{Ludwig} {et~al.}(2006){Ludwig}, {Allard}, \&
  {Hauschildt}}]{Ludwig2006}
{Ludwig}, H.-G., {Allard}, F., \& {Hauschildt}, P.~H. 2006, \aap, 459, 599

\bibitem[{{Luhman}(1999)}]{Luhman1999}
{Luhman}, K.~L. 1999, \apj, 525, 466

\bibitem[{{Luhman} {et~al.}(2003){Luhman}, {Stauffer}, {Muench}, {Rieke},
  {Lada}, {Bouvier}, \& {Lada}}]{Luhman2003}
{Luhman}, K.~L., {Stauffer}, J.~R., {Muench}, A.~A., {et~al.} 2003, \apj, 593,
  1093

\bibitem[{{Mart{\'{\i}}n} {et~al.}(2010){Mart{\'{\i}}n}, {Phan-Bao}, {Bessell},
  {Delfosse}, {Forveille}, {Magazz{\`u}}, {Reyl{\'e}}, {Bouy}, \&
  {Tata}}]{Martin2010}
{Mart{\'{\i}}n}, E.~L., {Phan-Bao}, N., {Bessell}, M., {et~al.} 2010, \aap,
  517, A53

\bibitem[{{Partridge} \& {Schwenke}(1997)}]{AMESH2O}
{Partridge}, H. \& {Schwenke}, D.~W. 1997, J. Comp. Phys., 106, 4618

\bibitem[{{Pettersen}(1980)}]{Pettersen1980}
{Pettersen}, B.~R. 1980, \aap, 82, 53

\bibitem[{{Phan-Bao} {et~al.}(2005){Phan-Bao}, {Mart{\'{\i}}n}, {Reyl{\'e}},
  {Forveille}, \& {Lim}}]{Phan-Bao2005}
{Phan-Bao}, N., {Mart{\'{\i}}n}, E.~L., {Reyl{\'e}}, C., {Forveille}, T., \&
  {Lim}, J. 2005, \aap, 439, L19

\bibitem[{{Plez}(1998)}]{Plez1998}
{Plez}, B. 1998, \aap, 337, 495

\bibitem[{{Rajpurohit} {et~al.}(2012){Rajpurohit}, {Reyl{\'e}}, {Schultheis},
  {Leinert}, {Allard}, {Homeier}, {Ratzka}, {Abraham}, {Moster}, {Witte}, \&
  {Ryde}}]{Rajpurohit2012a}
{Rajpurohit}, A.~S., {Reyl{\'e}}, C., {Schultheis}, M., {et~al.} 2012, \aap,
  545, A85

\bibitem[{{Reid} {et~al.}(2004){Reid}, {Cruz}, {Allen}, {Mungall}, {Kilkenny},
  {Liebert}, {Hawley}, {Fraser}, {Covey}, {Lowrance}, {Kirkpatrick}, \&
  {Burgasser}}]{Reid2004}
{Reid}, I.~N., {Cruz}, K.~L., {Allen}, P., {et~al.} 2004, \aj, 128, 463

\bibitem[{{Reid} {et~al.}(2007){Reid}, {Cruz}, \& {Allen}}]{Reid2007}
{Reid}, I.~N., {Cruz}, K.~L., \& {Allen}, P.~R. 2007, \aj, 133, 2825

\bibitem[{{Reid} \& {Gizis}(1997)}]{Reid1997}
{Reid}, I.~N. \& {Gizis}, J.~E. 1997, \aj, 114, 1992

\bibitem[{{Reid} \& {Gilmore}(1984)}]{Reid1984}
{Reid}, N. \& {Gilmore}, G. 1984, \mnras, 206, 19

\bibitem[{{Reiners}(2005)}]{Reiners2005}
{Reiners}, A. 2005, Astronomische Nachrichten, 326, 930

\bibitem[{{Reyl{\'e}} \& {Robin}(2004)}]{Reyle2004}
{Reyl{\'e}}, C. \& {Robin}, A.~C. 2004, \aap, 421, 643

\bibitem[{{Reyl{\'e}} {et~al.}(2002){Reyl{\'e}}, {Robin}, {Scholz}, \&
  {Irwin}}]{Reyle2002}
{Reyl{\'e}}, C., {Robin}, A.~C., {Scholz}, R.-D., \& {Irwin}, M. 2002, \aap,
  390, 491

\bibitem[{{Reyl{\'e}} {et~al.}(2006){Reyl{\'e}}, {Scholz}, {Schultheis},
  {Robin}, \& {Irwin}}]{Reyle2006}
{Reyl{\'e}}, C., {Scholz}, R.-D., {Schultheis}, M., {Robin}, A.~C., \& {Irwin},
  M. 2006, \mnras, 373, 705

\bibitem[{{Rossow}(1978)}]{Rossow1978}
{Rossow}, W.~B. 1978, \icarus, 36, 1

\bibitem[{{Rothman} {et~al.}(2009){Rothman}, {Gordon}, {Barbe}, {Benner},
  {Bernath}, {Birk}, {Boudon}, {Brown}, {Campargue}, {Champion}, {Chance},
  {Coudert}, {Dana}, {Devi}, {Fally}, {Flaud}, {Gamache}, {Goldman},
  {Jacquemart}, {Kleiner}, {Lacome}, {Lafferty}, {Mandin}, {Massie},
  {Mikhailenko}, {Miller}, {Moazzen-Ahmadi}, {Naumenko}, {Nikitin}, {Orphal},
  {Perevalov}, {Perrin}, {Predoi-Cross}, {Rinsland}, {Rotger}, {{\v S}ime{\v
  c}kov{\'a}}, {Smith}, {Sung}, {Tashkun}, {Tennyson}, {Toth}, {Vandaele}, \&
  {Vander Auwera}}]{HITRAN2008}
{Rothman}, L.~S., {Gordon}, I.~E., {Barbe}, A., {et~al.} 2009, \jqsrt, 110, 533

\bibitem[{{Skory} {et~al.}(2003){Skory}, {Weck}, {Stancil}, \&
  {Kirby}}]{Story2003}
{Skory}, S., {Weck}, P.~F., {Stancil}, P.~C., \& {Kirby}, K. 2003, \apjs, 148,
  599

\bibitem[{{Skrutskie} {et~al.}(2006){Skrutskie}, {Cutri}, {Stiening},
  {Weinberg}, {Schneider}, {Carpenter}, {Beichman}, {Capps}, {Chester},
  {Elias}, {Huchra}, {Liebert}, {Lonsdale}, {Monet}, {Price}, {Seitzer},
  {Jarrett}, {Kirkpatrick}, {Gizis}, {Howard}, {Evans}, {Fowler}, {Fullmer},
  {Hurt}, {Light}, {Kopan}, {Marsh}, {McCallon}, {Tam}, {Van Dyk}, \&
  {Wheelock}}]{Skrutskie2006}
{Skrutskie}, M.~F., {Cutri}, R.~M., {Stiening}, R., {et~al.} 2006, \aj, 131,
  1163

\bibitem[{{Tashkun} {et~al.}(2004){Tashkun}, {Perevalov}, {Teffo}, {Bykov},
  {Lavrentieva}, \& {Babikov}}]{Tashkun2004}
{Tashkun}, S.~A., {Perevalov}, V.~I., {Teffo}, J.-L., {et~al.} 2004, Proc.
  SPIE, 5311, 102

\bibitem[{{Testi}(2009)}]{Testi2009}
{Testi}, L. 2009, \aap, 503, 639

\bibitem[{{Tinney} {et~al.}(1993){Tinney}, {Mould}, \& {Reid}}]{Tinney1993}
{Tinney}, C.~G., {Mould}, J.~R., \& {Reid}, I.~N. 1993, \aj, 105, 1045

\bibitem[{{Tinney} \& {Reid}(1998)}]{Tinney1998}
{Tinney}, C.~G. \& {Reid}, I.~N. 1998, \mnras, 301, 1031

\bibitem[{{Tokunaga} \& {Kobayashi}(1999)}]{Tokunaga1999}
{Tokunaga}, A.~T. \& {Kobayashi}, N. 1999, \aj, 117, 1010

\bibitem[{{Tsuji} {et~al.}(1996{\natexlab{a}}){Tsuji}, {Ohnaka}, \&
  {Aoki}}]{Tsuji1996b}
{Tsuji}, T., {Ohnaka}, K., \& {Aoki}, W. 1996{\natexlab{a}}, \aap, 305, L1+

\bibitem[{{Tsuji} {et~al.}(1996{\natexlab{b}}){Tsuji}, {Ohnaka}, {Aoki}, \&
  {Nakajima}}]{Tsuji1996a}
{Tsuji}, T., {Ohnaka}, K., {Aoki}, W., \& {Nakajima}, T. 1996{\natexlab{b}},
  \aap, 308, L29

\bibitem[{{Udry} \& {Santos}(2007)}]{Udry2007}
{Udry}, S. \& {Santos}, N.~C. 2007, \araa, 45, 397

\bibitem[{{Unsold}(1968)}]{Unsold1968}
{Unsold}, A. 1968, {Physik der Sternatmospharen, MIT besonder Berucksichtigung
  der Sonne}

\bibitem[{{Valenti} \& {Piskunov}(1996)}]{Valenti1996}
{Valenti}, J.~A. \& {Piskunov}, N. 1996, \aaps, 118, 595

\bibitem[{{Veeder}(1974)}]{Veeder1974}
{Veeder}, G.~J. 1974, \aj, 79, 1056

\bibitem[{{Weck} {et~al.}(2003){Weck}, {Schweitzer}, {Stancil}, {Hauschildt},
  \& {Kirby}}]{Weck2003}
{Weck}, P.~F., {Schweitzer}, A., {Stancil}, P.~C., {Hauschildt}, P.~H., \&
  {Kirby}, K. 2003, \apj, 584, 459

\bibitem[{{Wing} \& {Rinsland}(1979)}]{Wing1979}
{Wing}, R.~F. \& {Rinsland}, C.~P. 1979, \aj, 84, 1235

\bibitem[{{Witte} {et~al.}(2011){Witte}, {Helling}, {Barman}, {Heidrich}, \&
  {Hauschildt}}]{Witte2011}
{Witte}, S., {Helling}, C., {Barman}, T., {Heidrich}, N., \& {Hauschildt},
  P.~H. 2011, \aap, 529, A44

\bibitem[{{Yurchenko} {et~al.}(2011){Yurchenko}, {Barber}, \&
  {Tennyson}}]{Yurchenko2011}
{Yurchenko}, S.~N., {Barber}, R.~J., \& {Tennyson}, J. 2011, \mnras, 413, 1828

\bibitem[{{Zacharias} {et~al.}(2005){Zacharias}, {Monet}, {Levine}, {Urban},
  {Gaume}, \& {Wycoff}}]{Zacharias2005}
{Zacharias}, N., {Monet}, D.~G., {Levine}, S.~E., {et~al.} 2005, VizieR Online
  Data Catalog, 1297, 0

\end{thebibliography}
\end{document}